# Web of Things and Trends in Agriculture: A Systematic Literature Review

Muhammad Shoaib Farooq, Shamyla Riaz, Atif Alvi

School of Systems and Technology, University of Management & Technology, Lahore, Punjab, 54770, Pakistan

Corresponding author: Muhammad Shoaib Farooq (e-mail: Shoaib.farooq@umt.edu.pk).

**ABSTRACT** In the past few years, the Web of Things (WOT) became a beneficial game-changing technology within the Agriculture domain as it introduces innovative and promising solutions to the Internet of Things (IoT) agricultural applications problems by providing its services. WOT provides the support for integration, interoperability for heterogeneous devices, infrastructures, platforms, and the emergence of various other technologies. The main aim of this study is about understanding and providing a growing and existing research content, issues, and directions for the future regarding WOT-based agriculture. Therefore, a systematic literature review (SLR) of research articles is presented by categorizing the selected studies published between 2010 and 2020 into the following categories: research type, approaches, and their application domains. Apart from reviewing the state-of-the-art articles on WOT solutions for the agriculture field, a taxonomy of WOT-base agriculture application domains has also been presented in this study. A model has also presented to show the picture of WOT based Smart Agriculture. Lastly, the findings of this SLR and the research gaps in terms of open issues have been presented to provide suggestions on possible future directions for the researchers for future research.

## I. INTRODUCTION

ADVANCEMENTS in technology not only changed the human lifestyle but also transfer traditional agriculture to Smart Agriculture (SA) and Smart Farming (SF) [1]. The Evolution of technologies made agriculture into the automatic and data-driven Smart Farming [2]. Another term that is being used for smart agriculture is Precision Agriculture defined for managing the soil, water, weather conditions through ICT technologies [3]. An approach was presented by Precision Agriculture Technologies (PATs) [4] for maximum production and for decreasing the harmful environmental causes. The key points related to the cost, long-term payback policy, and farms range by giving the training and technical support to the farmers [5]. Precision agriculture could be a processing cycle in which agricultural data is collected for analysis, evaluation to make better decisions in managing the fields that came under the umbrella of Smart Farming Technologies (SFT) [6] such as GNSS [7] and mapping technologies. The precision agriculture concept existed decades ago as a process or farmers' judgment. So this concept was named Precision Agriculture. A detailed technical explanation was also presented for agricultural industry [8].

The internet of things(IOT) also opened a new gate to the agricultural field by providing technological solutions in every agricultural domain [9]. The evolution of the Internet of things not only increased the number of devices but also generated a massive amount of data and estimated to be reached to 50 billion in 2020 [10]. Not only the devices, but the amount of data generated by them were also increasing. About 500 zettabytes of data were generated according to the estimation of CISO [11]. Due to the exponential increase in devices and data, the API amount also increased. This study introduced the model Web of Things, its main features, the SWT(Semantic Web Technology), and its usage in designing APIs [12].

The web of things made this increasing smart world of things more efficient and easy to use by engaging the web services into the development of these smart things. These smart things integrated the web within them and achieved interaction and communication over the web [13]. Since a massive amount of gadgets are connected to the Web, an unused Web of Things is developing, accompanied by virtual representations of physical or theoretical substances progressively open via Web advances [14].

A common concept of IoT, WOT, and SWOT is described in the figure 1:

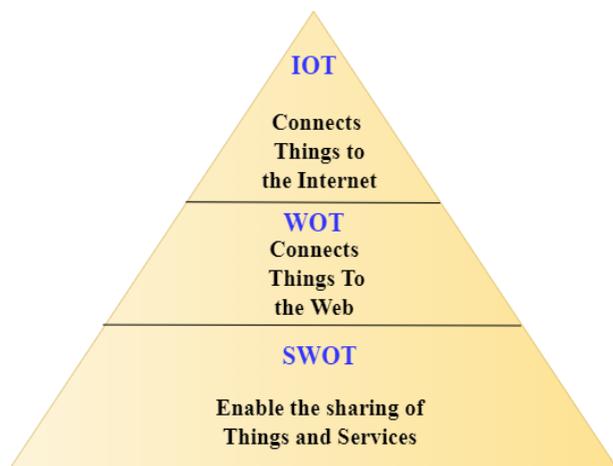

FIGURE 1: Basic concept of IOT, WOT, SWOT

Web of Things opened a new direction of research in agriculture for the reconstruction of agricultural land, alternative processes of cultivation based on modern market needs by introducing the decision making systems based on the web [15]. Study [16] presented a (VTEDS)-based structure for the advancement of smart sensor hubs with plug-and-play competencies. It was exceptionally quick, with intelligent sensors (found in Europe) able to self-record and self-configuration in a farther cloud (in South America) in less than 3s and to show information to remote users in less than 2s. Frameworks were also introduced for contribution in the advancement of the Internet of Things (IoT) to the Web of Things (WoT) [17]. As of late the ubiquity of utilizing the web of things in heterogeneous applications is expanded but various gadgets faced diverse encryption problems. To defeat difficulties, Semantic Web Technologies is viewed as the best arrangement [18].

**WOT integrated Model**:

The trends of introducing and emerging the various technologies in the Agricultural industry are growing rapidly. Many systems and frameworks, infrastructures have been presented which use IoT devices, wireless sensors [19], machines, and cameras for monitoring and controlling the crops' processes. There some infrastructures diagrams are described from Studies that presented the solutions by integrating the WOT services to enable the interoperability and communication between the devices and different layers of models, [20] [21] shown in figure 2:

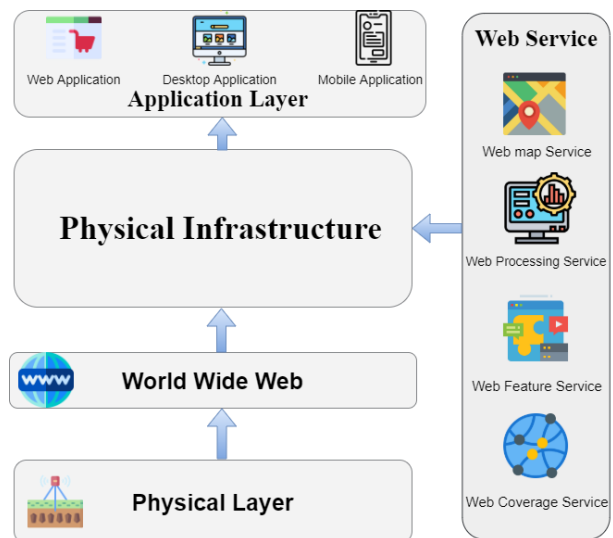

FIGURE 2: Open geospatial web physical model for PA

**WOT Communication Model**: Another infrastructure integrated the wot technology for connecting the IoT smart devices with the cloud for data sharing, storing, and communication purpose has shown in figure 3:

In recent years the Linked Open Data that integrated the different kinds of data made it useful for everyone in the world in every field [22]. Exponentially growing data led to the evolution of data and integration with the World Wide

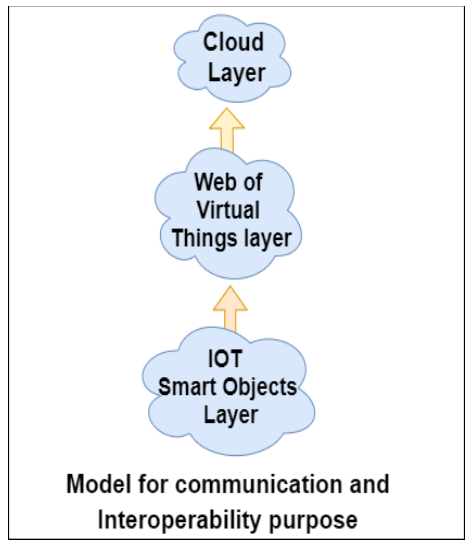

FIGURE 3: WoVT communication Model

Web as World Wide Semantic Web (WWSW) [23]. The interconnected world of people, devices, and services over the internet generated a lot of real-time data in different forms, codes, formats, massive information, and processes [24]. To store and access the real-time data in the desired formats, many algorithm techniques have been presented by integrating the existing and new one [25]. Semantic web [26] techniques made everything intelligent and efficient including the internet. MIMOS developed the platform named as a semantic technology platform that introduced the new features as well as improved the existed ones [27]. United Nation's FAO developed the largest agriculture ontology in the world as a multilingual agricultural vocabulary that provided many facilities and helped the organizations to publish their agricultural data and their models through AGROVOC [28]. Recently, wireless communication became a vital part of the connectivity and communication field with growing technology. Smart systems and sensors played the role of Controlling real-time processes easily and intelligently [29]. Smart systems, devices, and sensors stored and analyzed the data and also transferred traditional farming to smart farming. The accurate and secured data availability was overcome by Big data [30], AI, and cloud technology but there were some drawbacks of integrating these technologies, one of them was processing cost [31].

Another technology that integrated with IoT was Unmanned Aerial Vehicles (UAVs) for cultivating crops in smart agriculture [32]. With the enhanced technology, many agricultural processes could be monitored such as crop conditions and diagnosing the diseases [33]. By using the connectivity of different technologies in smart farming, the usage of fertilizers became less and effective with minimum cost [34]. Provided the 3D graphical representation of crops growing rate, the transparent supply of food items with the help of blockchain implementation. These emerging technologies can further enhance the precision agricultural field [35]. Blockchain technology became important in the field of precision agricultural applications. Another study was conducted on "Integrating blockchain and the internet of things in precision agriculture: Analysis, opportunities, and challenges" [36]. Developing a smart system to store, monitor, and analyze the data leads to the idea of developing a Blockchain-based system for precision agriculture. Blockchain changes the old methods to new methods with enhanced transparency, decentralization, and reliability [37]. Many frameworks have been introduced and implemented in agriculture for smart farming by many countries [38]. These models had the integration of IoT, Blockchain, WOT, Web-based messaging protocols, and other protocols such as smart application, MQTT, AMQP, DDS, REST HTTP, and web-socket protocol. These technologies and protocols were compared in their performance, functionality, efficiency and also proposed enhanced versions and combination of them for advanced Agricultural Industry [39].

The main goal of the study is to present the systematic literature review of the existing studies that defined the significance of WOT integration with agricultural applications for solving the agricultural industry problems related to their interoperability, handling, monitoring, and controlling the devices, applications, and other integrated technologies. The study investigates the State-of-the-art research conducted to solve the problems with the use of WOT. The novelty of this study is that it presented a taxonomy of the agricultural domains where WOT applications take place to solve the agriculture technological and technical problems.

Lastly, the remaining part presented the contribution of this paper in the form of different sections. Section 2 describes the WOT-agricultural field background. Section 3 presents the research methodology used for finding the relevant studies through defined research questions, criteria of inclusion and exclusion, and searching string to find the relevant research articles to the WOT- agricultural field. Section 4 defines the results of the elected studies gathered by extracting the data for this systematic literature review. Section 5 explains the discussion, proposed taxonomy, a wot based model of smart agriculture, open issues and challenges, research gap, and future directions for further research trends. Section 6 describes the threat to the validity of this study. The last section 7 presents the conclusion of the paper. Figure 4 shows the taxonomy of this paper.

## II. RELATED WORK

Web of Things has been a promising technology for solving several agricultural applications, systems integration, their handling, monitoring, and decision-making problems. WOT introduced remote sensors, data analyzing, and effective precision features in the existing agricultural frameworks, infrastructures, and systems. Several pieces of research have been conducted on WOT emergence in the agricultural technology sector to find the solutions to the arising challenges. Hence, few contributions have been made to evaluate the studies that presented the solutions to the agricultural problems. Some figures has showed the traditional and smart agriculture differences 5,6

Research related to the Web of Things was done on [20]. Wired and wireless [19] sensors played a vital role in Precision Agriculture for monitoring purposes. But the integration of divergent sensors into systems and their interoperability became an issue so a physical infrastructure with the integration of open geospatial service was proposed for processing, distributing, and integrating the monitoring details on World Wide Web. Web sharing service gives access to Precision Agricultural data widely. [40] Many web service architectures were proposed and designed but lack in sharing information and integration with diverse devices that lead to a Service Oriented Architecture (SOA) deployment for maintaining and updating purpose in PA(Precision Agriculture) systems that integrates the distributed flexible web services. Also, many studies have been done in solving the security issues of diverge networks, devices, and services but remain unsolved. In this study, many features of previous infrastructures were compared including SOA that was a lack in sensors support, process sharing, unknown models and have some complexity problems. So these all issues solutions and few more features were added in the proposed method.

The proposed SOA-based architecture were consists of four layers including physical, application, business, and Sensor layer. The physical layer contains devices and sensors for the transmission of environmental factor's measuring details. The application layer worked as an interface among systems and devices. An SOS sensing layer based on SWE used for distributing and storing the data over diverse sensing devices. The business layer is the higher-level layer for analyzing the agricultural processes and data. A primary monitoring pattern provides the data without communicating with sensors or sensor networks and users can access the details about agricultural measurements and much information on the Web. An advanced monitoring pattern provides the important data and process results by using SOS, WCS, WMS, and WFS services. This study proposed the sensor web idea and achieved the sensors, processes interoperability, and able to transfer, share, process, and integrate the data over World Wide Web. But there were still some gaps regarding multi-web services features that can be achieved in further work in the future.

A study [41] focused on the application of two technologies named WebGIS and IOT in Precision Agriculture. After analyzing the deployment, pros, and cons of PA in China, a Precision Agriculture Management System (PAMS) was deployed in a selected farm by integrating the WebGIS and IoT. It used the IoT for having perception accuracy ability and WebGIS processed the network geographical information inefficient way. This system had the following four modules: Agriculture management platform, spatial information infrastructure platform, and mobile client and IOT infrastructure platform. This system had the integration of several advanced techniques such as IoT, WebGIS, Internet and communication, Location-Based Service, GPS, RS to collect, transfer and publish the data that helped the users in monitoring and managing the production. PMAS has six components named as: the information infrastructure, database, local system, WebGIS, production management system, and mobile client. It suggested after analyzing the data to do the further process and saved time. By understanding the existing knowledge, experiences, and agricultural situations the system can be improved, new programs can be added and make it effective in the PA domain.

Further research was performed on "Agricultural Knowledge-Based Systems at the Age of Semantic Technologies" [42] Information can pave the way to new ideas, concepts related to the concerned disciplines. The organizations built knowledge-based systems that could help the researchers and general public to understand the in-formation, gaps, and limitations of their related disciplines and could provide new ideas and future work. Agricultural information and researches were available online but were not organized in a systematic form in systems and databases. The knowledge-based systems were used to investigate the neglected crops and utilized their information to make them useful. The database plan followed the standard vocabularies and ontologies that were developed a long time by the agrarian community and these ontologies helped in the development of semantic products to provide the direct question answers facility. CFF UCKB (Crops for the Future's Underutilised Crops Knowledge-Based System) provided the ground for agricultural researchers, organizations, and the public for educational and research purposes. The design contained the knowledge portal in the form of the Web interface, Knowledge database, and knowledge toolkit for mining and finding limitations through semantic products. The knowledge-based systems and tools played a vital role in storing, sharing, and exchanging information. In agriculture, this term referred to the document systems, knowledge networking system, social forum, world wide web, books. Agriinfo. is a web-based agricultural system developed by experts, AGRIS database agris.fao.org a document management system. A knowledge management system (KMS) was another technology that helped in finding the research gaps from existing papers and articles. Many decision and expert systems are classified into the knowledge-based systems that used the "if-then" query to provide the answers to the questions [43]. Several web-based systems used in agricultural field for different functions. [44], [33]. Another technique for providing the data online is Linked-Data that used the metadata standards for creating the templates to share and publish the data online. The system used the same standards to link the data and knowledge systems together for providing the datasets for research purposes. Social networking has started building the domain-specific feature for providing field-related information. The crops social network systems used this facility to linked the farmers, scientists, and researchers to communicate with each other to solves the problems. Another purpose of using the semantic technologies was to create a collaborative research environment (CRE) for underutilized crop issues. These were also used in making the documents and papers that could be readable to both machines and humans.

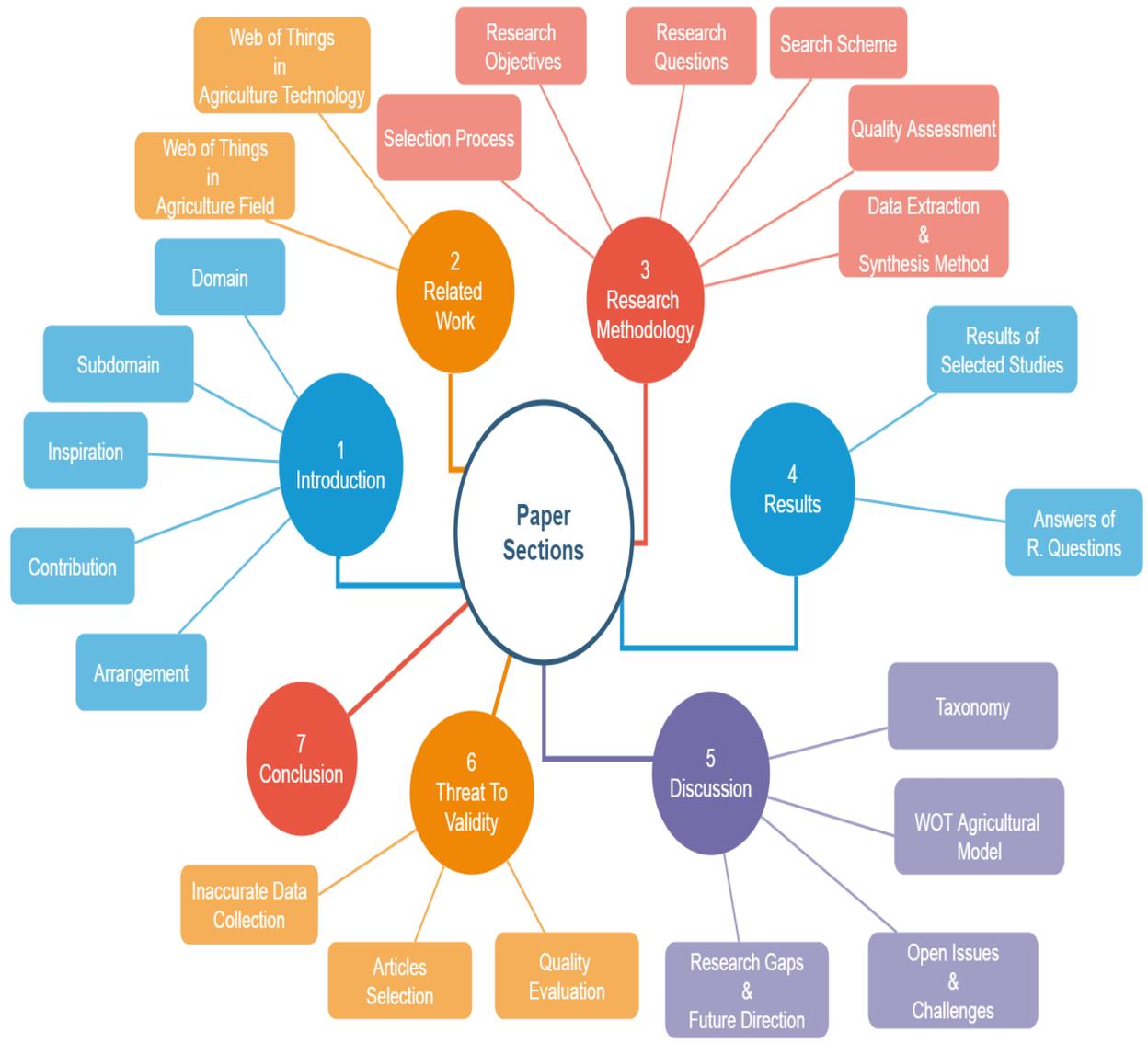

FIGURE 4: Paper Taxonomy

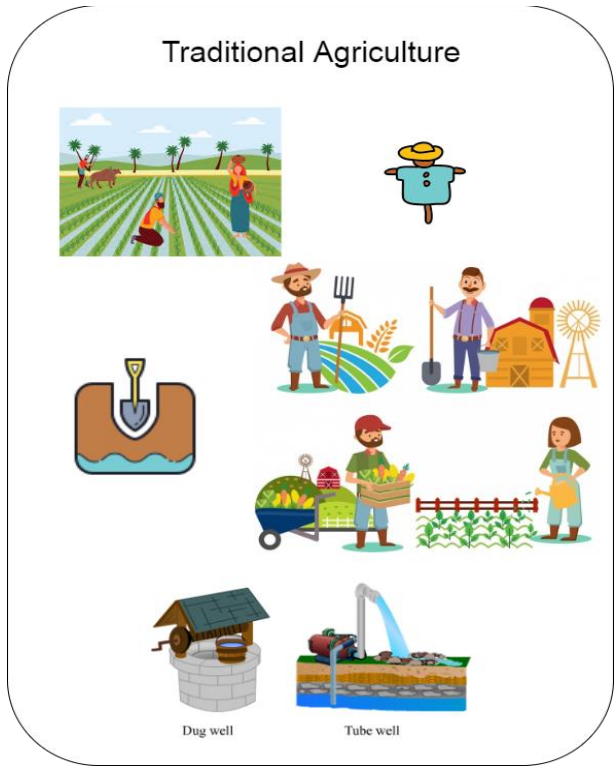

FIGURE 5: Traditional Agriculture

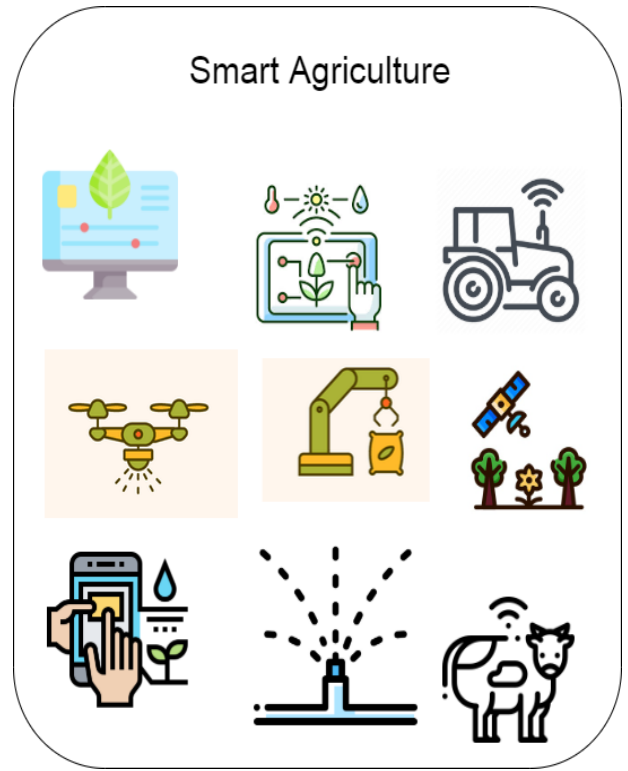

FIGURE 6: Smart Agriculture

Another study was conducted [45] in which another technique of using IT in the Integrated Pest Management was defined. It is a decision-based system that uses several methods and optimizes the control of all species of pests. The experts and researchers of the agricultural domain needed a tool for predicting and forecasting the pest according to the specified area. So a pest forecasting and information system with a web interface were developed that provided real-time functions. A software was developed in this work that used the World Wide Web technologies for storage and data mining purposes on weather data. It connected the NOA's network stations, downloaded the files with the .txt extension. In the next phase, this software was enhanced to the forecasting system that used the www and performed the real-time pest forecasting. The software development process was categorized into two stages: on the first stage, a customized WebGrabber obtained the weather data by open the NOA (National Observatory of Athens) meteorological stations network and the second one performed the real-time population calculation from the web on daily basis by using a web interface and stored information on MYSQL local server. An algorithm was developed for decision-making systems and forecasting. This blog's first step is BIOFIX that concerned with the evolution of pest species. The second step stored the data. The third one estimated the temperature. Then implied the condition that checks the temperature level. Then further proceed by calculating the input to the phenology equation and generated the results. This system was easily useable, less cost and simple architecture that virtually extendable.

In another study "A Web-Based IoT Solution for Monitoring Data Using MQTT Protocol" [46]. A web-based architecture was proposed to track, monitor, and analyze real-time agricultural data. The main aim was to monitor the various processes of IoT devices in the agricultural field. The sensors' data was stored and shared on the Web in the systems. The Web application architecture was a client-server three-tier platform with separate modules like graphical interface, storage section. A protocol named Message Queue Telemetry Transport (MQTT) was used for transferring the data from sensors and other monitoring devices to the users. The main feature of this protocol was having less power consumption, less complexity level, asynchronous communication. The generated and collected data was first sent to the servers and from then it was sent to the devices of users. But it still needs some progress for rural area networks.

A web of Things related work was done [47]. In agriculture providing the necessary amount of water to the crops for their better growth was also an important part. Weather conditions affected the water supply process so the farmers needed a system that also helped in making good and efficient decisions for this purpose. Real-time alerts and decision making needed to control the situations. In this regard, WSN (Wireless Sensor Networks) [19] played their part as an important and useful technology with the integration of IoT(Internet of Things) and WOT(Web of Things) for farmers to effectively and efficiently monitor and control the different situations related to the crops. The WOT provided the web services through with the farmers able to supervise the changes in the crops-related processes. Presented an alarm system for the estimation of water level needed, an application to manage the irrigation procedure, a graphical representation, real-time monitoring for soil situations, and an SMS alert service. This system can be further enhanced for efficient decision-making in the future.

Another study on the need of farmers about understanding the water standard for crops [48] presented that the detailed methods of agriculture caused a warning to the balanced ecosystems due to their poor water quality. This could help farmers to enhance their agricultural processes as they were familiar with the water supply. Instead of providing awareness to the farmers, the water systems were developed to aim at gaining the water supply according to the rules. A 1622WQ web-based application was developed and designed as user-centered for providing real-time information to the farmers. Some roadblocks such as (1) limited internet connection; (2) poor data quality; (3) operational issues were identified and taken care of in the application to provide good services. Data about rainfall parameters were also added. 1622WQ application prototype was developed with many versions using Shiny framework and an R package that were open source. The Interface contained three Tabs: "Locn" for showing all the locations to users and to filter them, "Map" for selecting the location, "Data" for showing the data of location. The web-based application was created by integrating the technologies and provided a single data and communication platform to farmers for different water supply projects and processes. This work highlighted the significance of collaboration and integration of heterogeneous approaches and technologies that led the world of traditional methods to the smart ones.

In an article, [49] research was conducted about improving the productivity of crops and farms. It was concluded that the production could be achieved by knowing the accurate, necessary weather and environmental conditions. This could be done by collecting data from agricultural studies about the quality of irrigation, farm's soil and weather conditions, etc. But the process of collecting the data was not efficient due to their different areas and was manual. For solving this problem the emergence of IoT and other technologies was used that included the different IoT devices, wireless sensors, and their networks, mobiles, smart systems, monitoring cameras, and weather stations. An IoT-based platform named SmartFarmNet was presented. It was developed by integrating the Semantic Web [26] Technology that enables the automatic collection of data about weather, soil, water conditions, filtered out the unnecessary data, precisions for better crop performance, and growth. It could virtually integrate the diverse devices and sensors and stored the data to perform analysis on them suggested the ideas. The main challenge in the development of the SFN platform was the management of a massive amount of diverse devices and networks. The solution to this challenge was designing a common API for all the integrated devices and sensors for representing their data through SWT. By integration of SWT, this platform was able to perform real-time analysis on data, extended its range to more domains, and could diagnose the errors and performance of devices. SWT helped to use the semantic web standards and enable the standards that were yet integrated into the existing systems. The evaluation and experiments were conducted by using the real farming data that validated the performance of the platform that ensured its scalable usage. This SmartFarmNet was the first platform in the world that supported a huge amount of diverse devices and networks with facilities for collecting, storing, analyzing the data, and forecasting the crop's performance.

One more framework [50] was projected with the integration of semantic web technologies for enhanced features to support the heterogeneous platforms, devices, sensors, and real-time data streaming in agriculture applications. The existing systems were improved in processing, analyzing the real-time data, decision making but they supported limited domains and platforms. The proposed framework has supported the integration of several diverse domains and platforms with the help of semantic web technologies. It provided a common platform with pipeline processing for applications, able to analyze the huge data, detected unknown events, and interoperated the heterogeneous devices, sensors, networks. Also supported online services like linked data, information, and open datasets on WWW. Evaluation of this framework showed good results of performance on the medium to large range farms. Also paved the new way for the integration of open standards and semantic web technologies in future agriculture.

A study was conducted on the adoption of WOT [51] stated that Applications of WOT required advancement in the solutions to its adaptation in common models for various goals. This study provided a solution for adaptation in the smart environment with the features of reusability and multi-purpose settings. Depended on the semantic technologies for the evaluation of information at run time, the solution was judged in the case of the Smart Agriculture framework that used the ASAWoO principles. Also discussed the process of designing models from traditional information sources. A component-based software named avatar was defined in the ASAWoO project for processing and controlling purposes. These avatars depend on the semantic architecture and could communicate with each other and high-level functions to involve WOT applications for similar objective achievement because the WOT applications had a ladder of functionalities. These composed functionalities are described in a detailed form in [52]. A multipurpose context adaptation process was provided that enable the WOT applications to answer the questions about domain-independent adaptations. A Smart agriculture framework was presented that demonstrates the proposed approach for achieving the accuracy and performance of implementation and evaluation. The outcomes validated the objectives about processing and question-answer adaptation. The proposed approach was also compared to identifying and managing related literature. Also, WOT application designer role was defined in collecting the contextual data and requirements of applications management workflow. In the end, some point of views about future work was defined related to the automatic generated domain-specific adaptation and modification rules.

In a previous research "IOT sys for Agriculture: web technologies in real-time with middleware paradigm" [53]. A specialized Middleware architecture was presented with the integration of real-time Web technology in the existing IoT systems for the urban area Agriculture. This model added the middleware level between the application and network layers. The web system in real-time consisted of many standards and protocols. The main function of integrating the web-based application was to store the data of every sensor device with a unique separate identifier. This web technology not only provided the storage for data but also presented the data in graphical visualization. Although this model was specialized web-based efficient architecture there is still a need of making it better to store, analyze and manipulate the massive amount of data generated in the future.

A study [27] was published that described the integration of Open Linked Data and Worldwide semantic web technology. In recent years the Linked Open Data that integrated the different kinds of data made it useful for everyone in the world in every field grew exponentially with heterogeneous formats and kinds that led to the evolution of data and integration with the World Wide Web as World Wide Semantic Web (WWSW). The Service-oriented techniques paved the way for the development of semantic technology to overcome many challenges related to performance, efficiency, and availability of various services over the internet world. World Wide Web Consortium standards opened the new world to the businesses, Agricultural Industry and attracted investors and entrepreneurs in every field. The interconnected world of people, devices, and services over the internet generated a lot of real-time data in different forms, codes, formats, massive information, and processes. Semantic web (kim2003)techniques made everything intelligent and efficient including the internet. Various organizations provided their data on the internet for experiments and analysis with existed methods, algorithms, and models. MIMOS developed the platform named as a semantic technology platform that introduced the new features as well as improved the existed ones. United Nation's FAO developed the largest agriculture ontology in the world as a multilingual agricultural vocabulary that provided many facilities and helped the organizations to publish their agricultural data and their models through AGROVOC. AGROVOC had 40,000 ideas in more than 20 languages related to the different fields such as Agriculture, fisheries, and forest, etc. It linked the other ontologies together such as corn, generic crop, chili, and tomato, etc. The four agricultural knowledge models were published in MIMOS. The machines and systems can use both ontologies by using linked data for searching the resources.

The research was performed on "A survey of semantic web technology for agriculture" [54] stated that the turning unstructured data into a useful form, Semantic web technologies have been playing a vital role. SWTs have also been used in solving the Agricultural domain problems as a supportive domain. Many big NGOs like FAO developed a huge amount of semantic technology resources for the Agricultural field. A survey was intentionally done to encourage future research on SWTs applications for solutions to the problems in agriculture. This survey provided a complete review of the already existed SWTs resources, their methodologies, standards of data interchange, and a survey on current SWTs applications. The articles for the survey were selected from peer-reviewed journals, books, and conferences for review. Processes in agriculture depended not only on crop types but also on other factors such as soil, water, weather, and environmental conditions and contained a massive amount of information that needed a common interlinked storage place. A common standard structure or platform for data storing and data integration could be achieved by semantic web technologies integration from real-time to non-real-time sources. The semantic resources for agriculture were reviewed in this study which consisted of taxonomies, controlled vocabularies, thesauri, and ontologies [55] that contains many agricultural sub-domains. Two approaches to creating the centered semantic resources for agriculture were described. One was to create a new one [56] and the other was to combine the existing resources [57]. The semantic web technology applications designed for the agricultural field were also discussed and considered in the study. As there was a huge amount of SWTs resources that were specified for the agricultural domain but few of them were used to solve the problems. Semantic web technologies seemed to be an important solution for solving problems in agriculture but fewer papers were found in the literature review related to SWTs in agriculture. Although there was not much literature review about this technology it can pave the way for future research work regarding SWTs applications in agriculture.

A study "ASPECTS OF WoT CONTRIBUTION TO SUSTAINABLE AGRICULTURAL PRODUCTION" [58] was done that performed analysis on the contribution of Web of Things in the agricultural domain. Following three steps were took to perform the analysis: 1. desires and estimates of analyst's region, 2. Movements of agricultural machines producers, 3. Status of existed accomplishments in the domain. The agricultural field moving to the industrial level, so the analysts forecasted the agriculture future. In attaining the key solutions agriculture faced two main challenges regarding weather changes and the long-term safety of surroundings. The changing frequency of IOT benefitted agriculture. The data from several internal and external sources and IoT devices were combined through WOT for achieving the solutions of various agricultural problems in making decisions for better productions. WOT and IOT are the main building blocks of smart agriculture and precision farming. Many US organizations made the attainments in the agriculture field such as AgJunction Inc., Monsanto and DuPont, Deere & Company, Raven Industries, and Trimble Navigation Ltd. A strong Encirca farm service with analytical architecture was launched by DuPont that used new technologies for transferring important data among farmers, presented a 3D sketch of every factor of farms for making further actions. (Poston, 2016). In Europe, The Future Internet Accelerator Programme for Internet-based innovation in the food and agribusiness provided 14 million euros for the growth of applications in this field. ( Verdouw et al., 2014). Another middleware project in the agricultural field was an open-source IoT Platform named Kaa (Popovic´ et al., 2016). Fujitsu and Microsoft offered a solution Plant Factory that was executed by Fujitsu in japan (Microsoft News Center, 2015) (Fujitsu, 2013). Robot technology also performed a huge role by introducing sensors, smart cultivating machines, and many monitoring and controlling devices that used the different technology integration for precision and decision making in agricultural farms. All the emerging technologies, protocols, smart systems made the smart farms for the future. This study concluded that the WOT is the main character for a smart farming future, so the farmers need to form plans for their high productivity and profit according to these technologies' ideas.

To provide the farmers with important information for better decision-making, IT played a vital role in the agricultural field. A research "Semantic Web-based Integrated Agriculture Information Framework" [59] was conducted that proposed a web-based knowledge model for gathering, mining, integrating the data from various sources to use for understanding and decision purposes in the agricultural field. The integrated Agriculture Information Framework (IAIF) model used web technologies to connect to the ontologies with field-related data and metadata for data extraction. A method was also provided for ontology to get the required data from different sources. IAIF also combined a decision-making module named Computer-based Discussion Support Systems (CBDSS) to help the farmers in better decisions ideas. The IAIF ontology contained three subgroups: Domain, Link, and Resource sections for providing the information of water, weather, soil, fertilizers for a huge production, profits to the farmers. Most of the agricultural data for decision purposes was stored on XML and Relational database and repositories, so the XML parser and D2R servers were used for knowledge gathering tasks. The link subsection of ontology linked the related databases and the Resource section represented the linking sources on the WWW. All of this was possible through web technologies to connect the field-related ideas with the databases for information and knowledge gathering. This research work was an addition to the existed work of providing the information to the farmers to make a better decision for huge production with less equipment and resources to fulfill the future needs. A discussion was also held to link the several diverse knowledge resources for enhancement in this work.

The involvement of the Web of Things fully changed the human's concept about wireless connection and device interconnectivity. Knowing and Understanding the worth of WOT, people utilized it by integrating it into the supply chain industry. Research is known as "Discussion on the Optimization of Web of Things Supply Chain of Agricultural Products and Information Sharing Based on RFID" [60] was conducted that introduced the WOT model and integration of it with the supply chain process to attain the sharing knowledge facility and to make SC effective over networks through RFID for maximizing the agricultural production. Various countries utilized the Web of things Supply Chain for agricultural food items qualitative and protection purposes. Integration technology of RFID introduced the real-time sharing facility for agricultural products information. In managing the production, producers entered the data into databases from which WOT connected the tags of RFID that contained the EPC identification codes of every agricultural food item. Everyone could know the food item's information through scanning these codes. For consumer's requirements, the supply chain process should be efficient and crystal clear for usage purposes that could be attained by WOT.

In recent years the IoT played a vital in many systems, devices, architectures, and applications but due to different protocols, networks, and object descriptions, it failed in the interoperable functions. A study "Towards An Interoperable Internet of Things Through a Web of virtual things at the Fog layer" [21] was conducted related to these integrations of IoT and WOT technologies. A lot of companies tried to develop the standards to solve the issues of diverse objects, platforms, and networks but this could not be accomplished. The already integrated method in the systems still did not support the IoT. To overcome these problems this study reviewed the several IoT infrastructures, different companies' projects, and applications of IoT and presented a conceptual framework. This consisted of three stages each for a different separate function. It contained a server named Web of Virtual Things (WoVT) that could be installed between the cloud and the middle layer named as Fog Layer of IoT to solve problems of compatibility. It also provided an interface REST for devices to incorporate at the lowest perception layer and with other servers of WoVT. A framework used for the semantic corporation model. The other intelligent devices do not need to hold up for requests or to answer any query. It can express the received messages with a consistent style in a non-realistic device. This non-realistic device act as the realistic one. It answered the queries when the actual device is not in the active position and when they become active it updated them regarding queries. To check the power of working, capacity and safety measures, the examination and valuation task was performed. The simulation process outcomes showed that the emergence of WOT with IoT can be highly useful and successful. For the enhancement of secure working of virtual things, more methods needed to be employed.

Another research [61] was performed on the contribution of WOT in the agricultural domain. The integration of IoT into the sensors and devices for environment monitoring enhanced the agriculture field by introducing the decision-taking feature and turned it into intelligent agriculture. Thus enhancement also led to some issues about obstruction and revealing the agricultural data at the time of gathering data. To overcome this issue, this study presented a framework named "ASAWoO" that used some obstruction control techniques that could be deployed at the middle level in IoT architecture that enabled the device controlling with the help of WOT standards and codding rules and computing techniques for obtaining the interconnectivity among various devices and sensors. WOT helped the devices and sensors to communicate and interact with each other for performing the tasks. This model showed the devices virtually over the web to enable the users to communicate with devices and to control and monitor them easily. An experiment was conducted for four months on a farm to judge the effectiveness of this presented framework. The output of this deployment results that the emerging IoT, WOT, and opportunistic computing that could give great success to the agricultural field. It was also realized that this proposed model can also be integrated into other fields and provide the best functioning to them.

A research "Publishing Danish Agricultural Government Data as Semantic Web Data" [62] conducted in providing and getting the agricultural data freely and easily. Recently the well-organized data became available online for free use. All this was happened due to the rapidly growing semantic web and linked open data technologies. These technologies emergence made the data available for public use with free of cost facility. Many countries' governments used these technologies and provided free data for public use. The well-known states UK and USA also utilized the new trends and provided useful free data in many fields such as in health, educational and agricultural domains for maximizing the economy. The integration of web technology with Linked Open Data not only enabled the data available freely but also provided the facility to connect different sources and gathered the data easily from them. But this data most of the time was not in the furnished form that made it impossible to use it for analysis and proceeding purposes and also made it hard to connect with other databases. To solve this problem, this study proposed a method of gathering and proceeding the data and connecting with different databases by using a web framework named "RDF" for the Danish government in the Danish language. Also enabled the feature of connecting the agricultural data to the organizations and made them find the answers related to the questions that were hard to do. This model procedure gets the data, performs analysis, cleans the data, and then linked it with other data sources. In the end, the model was judged by experimenting that showed good results.

An article was published about the management of historical property [63]. During the latest years, administrations, communal institutes, and neighborhood groups have dedicated developing interest to the identity of favorable techniques for the upkeep and beneficiation of cultural historical past assets. Plans at the control of cultural background property are primarily based totally on multiple, frequently disagreeing, standards and on many risks, and doubtlessly non-mutual characters and participants. Regarding this, the give and take among the renovation of property ancient representative morals and the edition to opportunity and carefully worthwhile makes use of play a crucial task in funding choices, multi-standards examines offer sturdy academic and procedural structures to help to decide entities about draft and employment of flexible reusable techniques for cultural history and civic actual property. A multi-standards selection method was offered for rating beneficiation techniques of history culture belongings aimed toward selling their recovery and preservation, in addition to growing cultural and financial welfares. A singular software of the A'WOT for evaluation in assisting the plan and employment of control techniques of abandoned background properties of culture. This application contains two methods AHP and SWOT that helped in making decisions, managing resources of nature and tourism, and finding problems and their solutions. For effective results of this methodology, all the methods should be integrated and processing stages should be arranged in proper step-wise manners.

Previously a lot of ideas were presented for creating a distant study stage with the facility of practical teaching to the students of four study areas: Science, Engineering, Mathematics, and Technology. A research was conducted on "Contribution to the Setting up of a Remote Practical Work Platform for STEM: The Case of Agriculture" [64] stated that the surveys of natural and life sciences study areas showed that the subjects given to the students were not completely

VOLUME 1, 2020　　　　　　　　　　　　　　　　　　　　　　　　　　　　　　　　　　　　　　　　　　　　　　　　　　　　　　　　　　　　11

well-teaches. A practical part should be added for better study but the visit to the biological divergent region could be dangerous for persons. This study also took part in the enhancement of distance learning to provide a virtual visiting place and sharing different facilities by joining the features of WOT and WebRTC a multimedia server technologies. The focus of work was primarily for the agricultural field but the experimental outcomes could also be related to other field areas. This presented answer provides the facility by which some learners and teachers could go to the field trips while other learners could be facilitated by a live broadcast of that trip and get the experience. The key point of using WOT was the easy establishment with APIs and standards that enabled the different objects to interact with each other by web language. Small level servers could also use in the devices with their rules. This framework consists of three stages name IoT, API, and Web Application interface. By implying this framework a communication among different users could be enabled on the simple web browser and get the data from already deployed devices. This proposed framework could be deployed and enhanced agricultural distant learning.

From the past few years, with the enhancement of technologies, the internet, communication networks, and devices, the world has become a global village and things became smart that is called the Internet of Things today. A study "Smart Sensors from Ground to Cloud and Web Intelligence" [65] was done on Embedded Intelligence (EI) an integrating research field that works on disclosing the actions of every single person, things, multidimensional shapes and also find out the hidden marks of intelligent devices. The data extraction was performed by using open-source data in the agricultural field to create useful ontologies and information resources for making effective decisions. The previous agricultural data was analyzed smartly to create the new material for enhancement purposes. The previous research record, properties, applications, frameworks, and EI research-related gaps were reviewed in this study and demonstrated an intelligent water supply application. The main aim was to simplify the threats and new directions of Intelligent Building Technologies and to make smart managing systems for agricultural processes. The semantic web technologies were used for the enhanced searching process, publishing the data online in well-organized form and connecting the devices for communication and working. The intelligent agricultural knowledge should be analyzed by scientist's teams to check the reliability to deploy in actual farms and agricultural areas by the countries for effective production. International trusted standards are also needed for secure data availability.

Even though many studies suggested the WOT and other technologies emergence and applications into the agricultural domain to overcome the interoperability, performance, production, management challenges. Very little literature work is found on it and no study presented any taxonomy for clear explanation and future opportunities for further research in this regard. However, this work presented a systematic literature review of recent relevant studies that were conducted to provide the solutions to the agricultural problems. The novelty of this study is that we have presented a taxonomy of WOT applications to the agriculture domain and a wot based agriculture model. In last, open issues and paths to future research have been pointed out.

## III. RESEARCH METHODOLOGY

The proposed methodology will be a Systematic Literature Review that will be conducted to collect, study and categorize the existing and ongoing researches and the protocols, standards, emerging technologies in them [66]. Figure 7 shows the process of conducting the systematic study. The process consists of three steps: Study Preparation, Study Conduction, Analysis, and Results. The selected papers will be classified according to their research topics, research types, research approaches, and application domains. The systematic study papers will be extracted and categorized according to their approaches. This SLR will be discussed and will get suggestions from researchers for future directions [67]. The followed methodology has been proposed by [66].

### A. RESEARCH OBJECTIVES

The main objectives related to the selected study are following:

- To provide a sketch of continuing research about WOT technologies within the Agricultural industry.
- To recognize and unveil the existing work, limitations, and future work related to WOT in Agriculture applications.
- To present a summary of research movements.
- To identify field-related publications of the study.

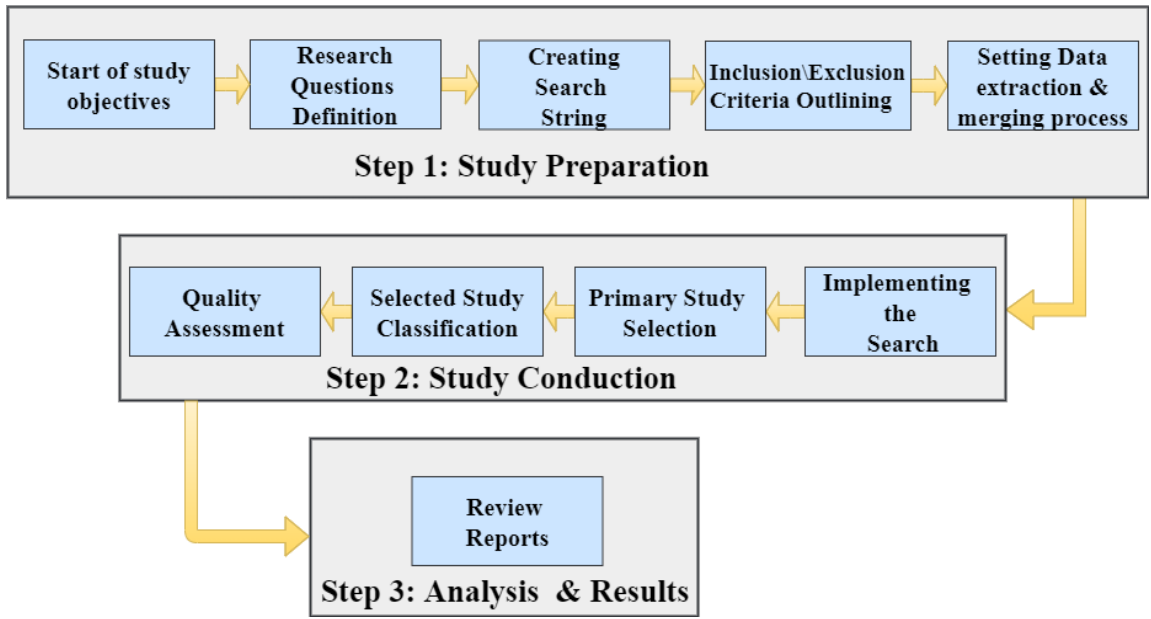
FIGURE 7: Systematic Literature Review Process

## B. RESEARCH QUESTIONS

For getting a complete picture of the elected study, five research questions (RQs) have been defined with the related motivations for the selected study. Table 1 given below show all the questions. Answers to given questions will help in identifying and classifying the current research trends, their drawbacks, and future directions related to WOT Agricultural.

## C. SEARCH SCHEME

A search string has been used to collect the related articles on the research topic. The search strings used for conducting the automatic search in the Scientific databases are given in the figure 8 and the search scheme to get the most relevant papers from scientific databases has shown in figure 9.

FIGURE 8: Search string used in scientific databases

We conducted the search using specific keywords and used only the applications in the search string for agriculture. We have also used the WOT sensors/devices, services and technology in agriculture in the search string. Given the research and knowledge-based databases were utilized for obtaining the related work for this study: Elsevier, IEEE Digital Library, MDPI, Science Direct, Google Scholar, and Springer. Google Scholar was used to getting the bibliometric studies.

TABLE 1: Research Questions

| No. | Research question | Main motivation |
|---|---|---|
| RQ1 | What are the major targeted publication channels for WOT Agricultural research? | To classify where the WOT Agricultural research can be found as well as a reliable publication source for future study. |
| RQ2 | What is the changing frequency of approaches related to WOT agricultural studies/researches over time? | to classify the publications shifting overtime of WOT-Agriculture Research. |
| RQ3 | What are the research types of WOT Agricultural study? | to find out the research approaches concerning WOT in the study. |
| RQ4 | What are the current research topics and gaps in WOT agricultural research? | to understand the current research topics that will help in finding the future directions for research in this field and to acknowledge the unanswered questions in existing agricultural research. |
| RQ5 | What are the approaches that were presented to address the problems in WOT agricultural research? | To determine the existing approaches presented in the existing WOT agricultural study. |

### D. STUDY SELECTION PROCESS

The selection process focused on identifying the best-related research articles for the selected study. The articles that appeared in more than one source were considered only once according to the research order. We have evaluated every paper by examining its Title, keywords, and abstract carefully, in regards to whether it should be included or excluded. In the second phase, the identical titles and titles not related to the selected study were removed. In the second phase, the papers were included/exclude according to the inclusion/exclusion criteria explained in the table 2

TABLE 2: Inclusion/Exclusion criteria applied for the elected study

| Inclusion criteria | Exclusion criteria |
| --- | --- |
| IC1- Articles presenting concepts and integration of WOT | EC1- Articles that are not focused on WOT applications |
| IC2- Articles that are focused on WOT applications and their implementation | EC2- Articles not presenting new and emerging ideas |
| IC3- Articles presenting wot problems/goals | EC3- Articles presenting general focus on WOT application integration based model |
| IC4- Articles presenting WOT standards/protocols/tools | EC4- articles not related to the search string |
| IC5- Studies published in English Language | EC5- Articles that are published before 2010 |

The result of the selection process demonstrated in the figure 10.

### E. QUALITY ASSESSMENT

A crucial phase in reviewing the method is to achieve the quality of elected papers. Generally, the Quality Assessment(QA) is carried out in Systematic Mapping Study and Systematic Literature Review. A questionnaire was prepared for executed the quality of selected papers for this study by following an SLR. [68]

**(1)** The study contributes to the WOT in Agriculture. Possible responses were: Yes (+1), Moderate (+0.5), and No (+0).

**(2)** The Study represented a clear solution for agricultural field through WOT. Yes (+1), Moderate (+0.5), and No (+0).

**(3)** Limitations and future directions of the study are clearly described. Yes (+1), Moderate (+0.5), and No (+0).

**(4)** The paper was published in a renowned and trustworthy publishing channel. Considering the following journal rankings and conference ranking (CORE) web-sites: https://www.scimagojr.com/index.php (Q1, Q2, Q3, and Q4), http://www.conferenceranks.com/ (A, B, and C) and the Citation Reports for journal ranking (JCR), this question was evaluated. A possible answer to this question for Conferences and Journals Rankings:

- If ranking is CORE A (+1.5),
- If ranking is CORE B (+1),
- If ranking is CORE C (+0.5),
- If it's not ranked in CORE (+0)

Journals ranking:

- Where ranking is Q1 (+2)
- Where ranking is Q2 (+1.5)
- Where ranking is Q3 or Q4 (+1)
- if not ranked in JCR list (+0)

A final score of each study (that ranges from 0-5) is provided by estimating the total score for each question.

### F. DATA EXTRACTION APPROACH AND SYNTHESIS METHOD.

The data extraction approach aimed to provide reliable answers to the research questions(RQs).

**RQ1**. To answer this research question (RQ), a publishing channel as well as a means of the articles is needed to determine.

**RQ2**. Papers should be arranged according to the publishing year to identify the frequency of publication trends.

**RQ3**. A research type can be specified by the following classification [69]:

- Solution proposal: Solving methods for agriculture problems are offered. the solution can be novel or an evolution of a recognized method. The potential advantage and significance of the solution are shown by justification or with few examples.
- Conceptual Research: These studies simplified the concepts by perceiving and studying the existing information on the WOT applications. Any practical experiments are not involved in this.
- Evaluation Research: Analysis and Valuation to WOT-Agriculture system/approach are carried out. Also, it

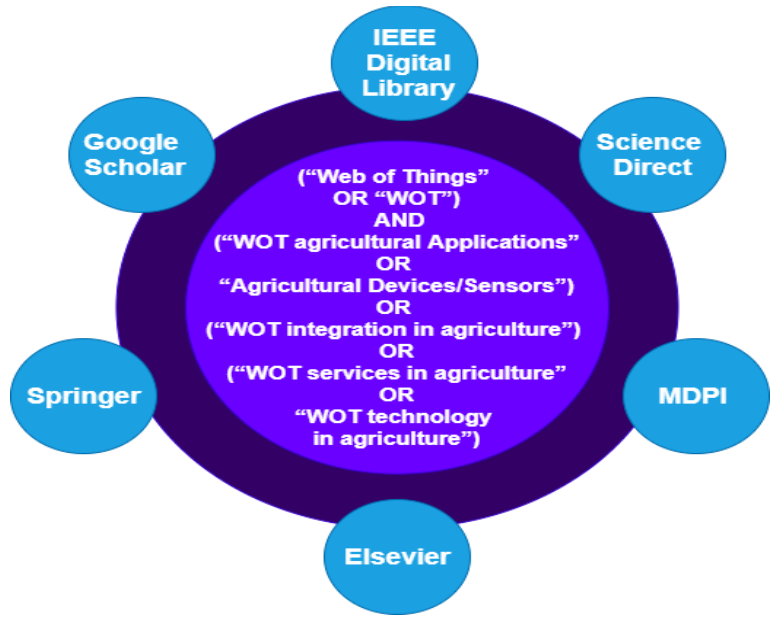

FIGURE 9: Search scheme used in scientific databases

concerns realizing issues in WOT-Agriculture applications.
- Others: such as surveys, system, architecture, Development, experimental, reviews, performance analysis, and models.

**RQ4**. Understand the existing research trends about Web of Things (WOT) Agricultural solutions is the dominant research query of study. We might be able to understand the study on WOT-Agricultural applications by collecting the targeted articles from scientific journals, finding the research gaps, and able to present research trends. This SLR (systematic literature review) will facilitate general practitioners and new researchers to enhance the knowledge on existing research matter to do further work on WOT-agricultural solutions.

**RQ5**. An approach can be classified according to [69], in the following types:

- System: It can be a proposal that provided monitoring solutions, controlling, and decision making in agriculture By using WOT.
- Framework: A specific theoretical or conceptual framework formed for supporting or attaining the interoperability and optimization of IoT systems through WOT.
- Application: A proposed solution designed to monitor, manage the agricultural processes and for providing awareness to the farmers or users through Web-based applications.
- Method: A structure suggested and a step-by-step procedure for obtaining agricultural knowledge through WOT services.
- Infrastructure: Supervisory Physical system needed regarding agricultural monitoring or precision purpose through WOT services.
- Architecture: A method, plan, or design about agricultural sector techniques.
- Guideline: An example of a practice or discussion that can be used to find solutions for agriculture by WOT technologies.
- others: analysis, platforms, etc.

The process's main aim is about the fundamental researches that are classified according to every question of research, bringing up the fundamental researches along their rank based according to and presenting a graphical view of outcome of classification.

## IV. RESULTS

This section is about the ending results of the research questions stated in the Table by explaining in detail. We selected the studies by screening process and added them in contribution to the WOT Agriculture domain based on their significance value.

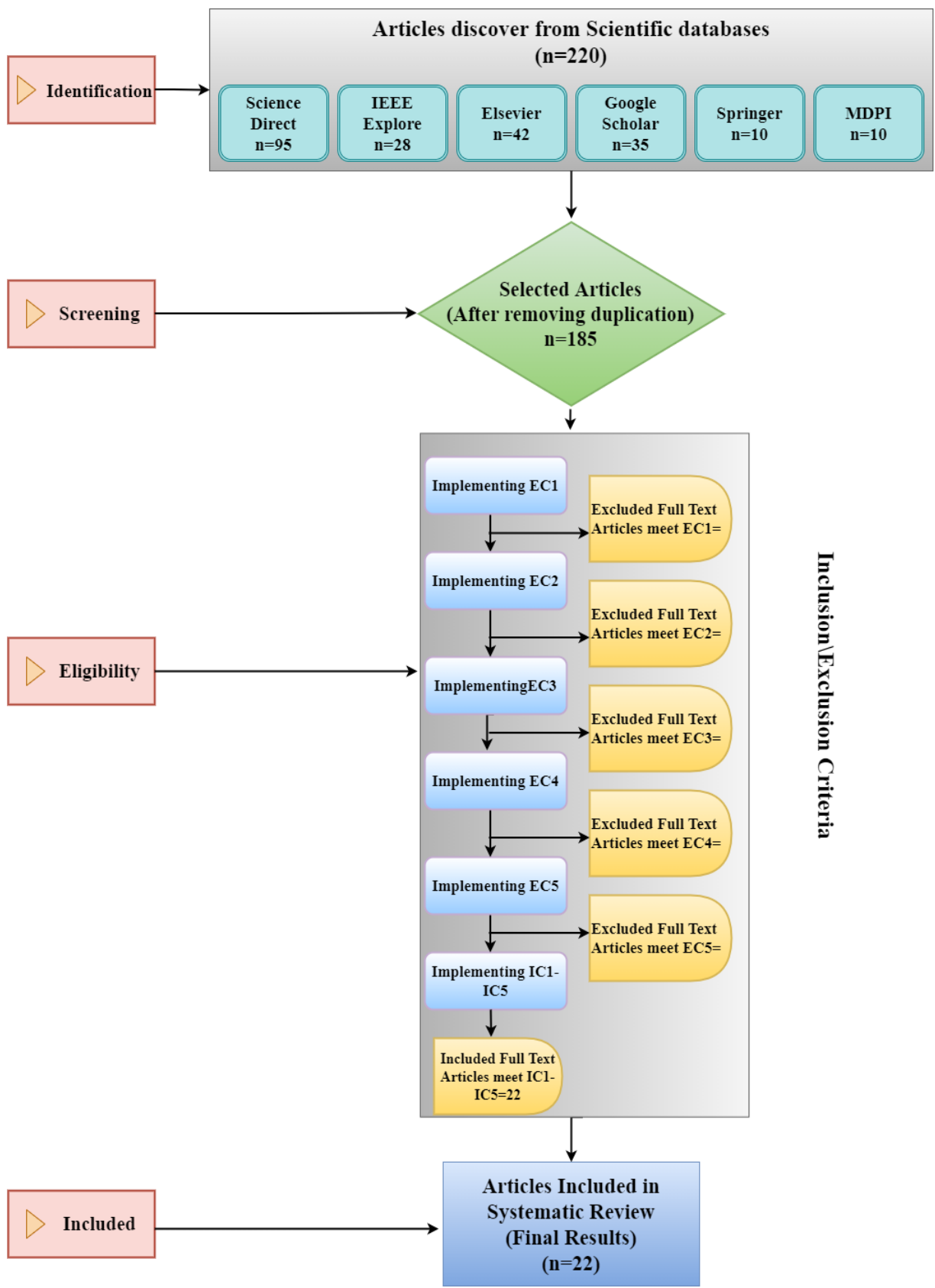

FIGURE 10: Study selection process

## A. SELECTED STUDIES RESULTS

Entirely 220 research studies were checked thoroughly in terms of their titles, keywords, and abstracts from which 198 articles were excluded and 22 articles were included according to criteria. The 22 papers were studied carefully for answering the research questions of this study. The selected papers with the details of their classified results showed in the Table 4.

## B. RQ1. WHAT ARE MAJOR TARGETED PUBLICATION CHANNELS FOR WOT AGRICULTURAL RESEARCH?

The journals and conferences were the only Publication channels involved in this Systematic Literature Review. Figure 11 showed the publication channels of the selected papers.

All papers were published in the journal (12)(0.54%) and conference (10)(0.45%). Besides, all the publication sources where the elected papers were published are listed in the table 3. All the selected papers used different publication sources.

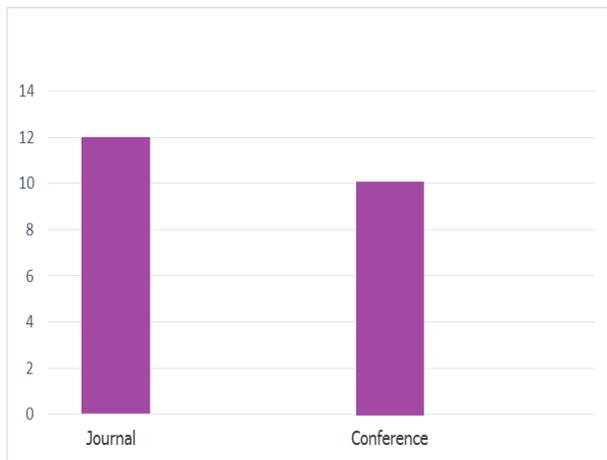

FIGURE 11: Publication channel

## C. RQ2. WHAT IS THE CHANGING FREQUENCY OF APPROACHES RELATED TO WOT AGRICULTURAL STUDIES/RESEARCHES OVER TIME?

The papers selected for this study were published in the years from 2010 to 2020. Types of publications over years have shown in figure 12 . The yearly trends of WOT integration in agriculture have shown in the figure 14.

TABLE 3: Publication Channels of the elected studies

| source of Publication | References |
|---|---|
| Computers and Electronics in Agriculture | [20] |
| 2013 21st International Conference on Geoinformatics | [41] |
| International Journal of Knowledge Engineering | [70] |
| Integrated protection of fruit crops IOBC-WPRS Bulletin | [71] |
| 2016 International Conference on Smart Systems and Technologies (SST) | [72] |
| The 12th International Conference on Future Networks and Communications | [73] |
| Journal of Environmental Modelling and Software | [74] |
| SENSORS | [75] |
| 2016 IEEE 3rd World Forum on Internet of Things (WF-IoT) | [76] |
| International and Interdisciplinary Conference on Modeling and Using Context | [77] |
| 2018 IEEE International Autumn Meeting on Power, Electronics and Computing | [78] |
| INFORMATION PROCESSING IN AGRICULTURE | [79] |
| Journal of Integrative Agriculture | [27] |
| Analele Universității din Oradea, Fascicula: Protecția Mediului 2016 | [80] |
| 2010 Second International Conference on Computer Research and Development | [81] |
| Journal of Physics: Conference Series (JPCS) | [82] |
| Future Generation Computer Systems | [83] |
| Future Internet | [84] |
| JIST 2014: Joint International Semantic Technology Conference | [85] |

Most publishing year of papers was 2016 because that was the growing era of WOT emergence in every domain well as in the agriculture domain with the emergence of IoT

and other technologies. The 2nd most publishing years were 2018 and 2019 in which most researches were con-

| | |
|---|---|
| Sustainability | [86] |
| International Conference on e-Infrastructure and e-Services for Developing Countries | [87] |
| IFAC-PapersOnLine | [88] |

ceptual, proposed, and contributions about WOT integration with internet of things systems and architectures. As shown in the graph that no paper was published in 2011 and most of the articles were published in journals. WOT integrated into many ways in the agricultural field and many of them were frameworks, systems, and web applications. As this study was conducting in the year 2020 so it is not likely to unveil the exact amount of studies published in 2020. The 2 to 3 papers were published in journals as well as in conferences every year.

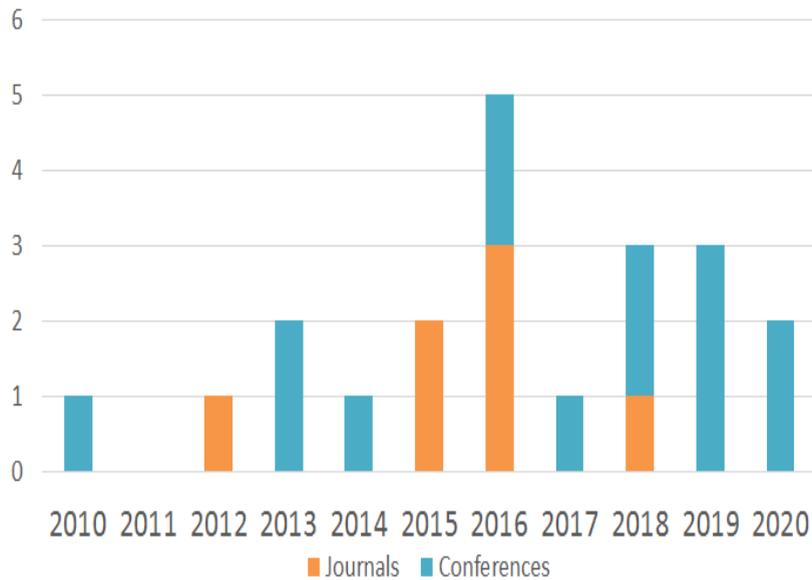

FIGURE 12: Publications over Years

### D. RQ3. WHAT ARE THE RESEARCH TYPES OF WOT AGRICULTURAL STUDY?

In this systematic study, nine (9) types of research was recognized in selected 22 articles that included : Solution proposal (6 studies)(0.27%), Experimental research (3 studies)(0.14%), Proposed System (3 articles)(0.14%), Proposed Model (2 articles)(0.09%), Contributional Research (3 articles)(0.14%), Conceptual Research (2 articles)(0.09%) while just 1 study was Implementation and Evaluation Research, a Survey and an Evaluation Research. Results of these types shown in the figure 13 and all the types shown in figure 15. The given graphical representation is showing the elected studies (Solution proposal) are mostly the solution for the Agricultural problems. Some Experimental and implementation & evaluation of WOT-Agricultural solutions were also introduced.

[20] In this study, the author proposed the Service Oriented Architecture (SOA) as a solution to integration and interoperability of sensors into the systems for monitoring and sharing purposes by combining the previous infrastructure features and new features. The new infrastructure contained the sensor web idea and web services for sharing, distributing, processing the agricultural data over the World Wide Web. In this study [46], the author found the difficulties in getting and monitoring the real-time data and proposed a web-based infrastructure with the use of MQTT protocol

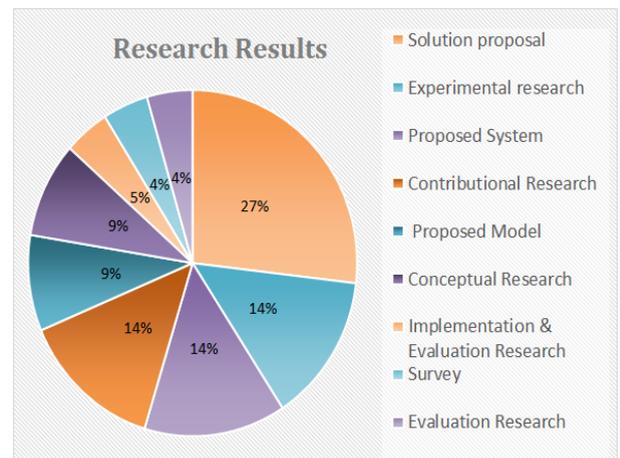

FIGURE 13: Research Results

for monitoring and tracking the devices as well as analyzing, storing, and sharing the real-time data to the users. An author examined [48] the water supply systems problems regarding the awareness to the farmers and presented a single web-based application platform for farmers to get real-time information about water quality. This application also resolved the issues of internet connection and data quality.

TABLE 4: Classification of the selected studies

| Ref. | P.Year | P.Channel | Research Topic | Research Type | Research Approach | Application Domains | (1) | (2) | (3) | (4) | T.score |
|---|---|---|---|---|---|---|---|---|---|---|---|
| [20] | 2015 | Journal | Integrated open geospatial web service enabled cyber-physical information infrastructure for precision agriculture monitoring | Experimental Research | Infrastructure | Monitoring | 1 | 1 | 1 | 2 | 5 |
| [41] | 2013 | Conference | A Precision Agriculture Management System Based on Internet of Things and WebGIS | Solution proposal | Management system | Monitoring and Management | 1 | 1 | 1 | 0 | 3 |
| [42] | 2015 | journal | Agricultural Knowledge-Based Systems at the Age of Semantic Technologies | Solution Proposal | Database system | Storing and Analysis | 0.5 | 0.5 | 1 | 1 | 3 |
| [45] | 2013 | Conference | Real time pest modeling through the World Wide Web:decision making from theory to praxis | Proposed System | Decision making and Forecasting system | Storing and Analysis | 1 | 1 | 1 | 0 | 3 |
| [46] | 2016 | Conference | A Web-Based IoT Solution for Monitoring Data Using MQTT Protocol | Proposed Model | Web Application | Monitoring | 1 | 1 | 0.5 | 1.5 | 4 |
| [47] | 2017 | Conference | Monitoring system using web of things in precision agriculture | Proposed Model | Monitoring System | Monitoring, Controlling and Decision making | 1 | 1 | 1 | 0 | 3 |
| [48] | 2016 | Journal | 1622WQ: A web-based application to increase farmer awareness of the impact of agriculture on water quality | Proposed Solution | Web Application | Monitoring | 1 | 1 | 0.5 | 2 | 4.5 |
| [49] | 2016 | Journal | Internet of Things Platform for Smart Farming: Experiences and Lessons Learnt | Implementation and Evaluation Research | Platform | Analysis | 1 | 1 | 0.5 | 2 | 4.5 |

TABLE 4: Continued Classification of the selected studies

| Ref. | P.year | P.Channel | Research Topic | Research type | Research Approach | Application Domains | (1) | (2) | (3) | (4) | T.score |
|---|---|---|---|---|---|---|---|---|---|---|---|
| | | | | Classification | | | | Quality Assessment | | | |
| [50] | 2016 | Conference | Agri-IoT: A Semantic Framework for Internet of Things-enabled Smart Farming Applications | Solution Proposal | Framework | Analysis | 0.5 | 0.5 | 1 | 0 | 2 |
| [51] | 2017 | Conference | Multi-purpose Adaptation in the Web of Things | Experimental Research | Platform | Storing and Analysis | 1 | 1 | 1 | 0.5 | 3.5 |
| [53] | 2018 | Conference | IOT sys for Agriculture: web technologies in real time with middleware paradigm | Proposed System | Architecture | Storing and Analysis | 1 | 1 | 1 | 0 | 3 |
| [54] | 2019 | Journal | A survey of semantic web technology for agriculture | Survey | Guideline | Analysis | 1 | 0.5 | 1 | 2 | 4.5 |
| [27] | 2012 | Journal | World-Wide Semantic Web of Agriculture Knowledge | Evaluation Research | Architecture | Analysis | 1 | 1 | 0.5 | 2 | 4.5 |
| [58] | 2016 | Journal | ASPECTS OF WoT CONTRIBUTION TO SUSTAINABLE AGRICULTURAL PRODUCTION | Contributional Research | Existing Aspects Analysis | Analysis | 1 | 1 | 1 | 1 | 4 |
| [59] | 2010 | Conference | Semantic Web based Integrated Agriculture Information Framework | solution Proposal | F I R | | | | | | |
| [60] | 2020 | Journal | Discussion on the Optimization of Web of Things Supply Chain of Agricultural Products and Information Sharing Based on | Conceptual Research | | | | | | | |

~~Framework Integration 1 1 1 0 3~~

Guideline Optimization 1 1 1 0 3

TABLE 4: Continued Classification of the selected studies

| Ref. | P.Year | P.Channel | Research Topic | Research Type | Research Approach | Application Domains | (1) | (2) | (3) | (4) | T.Score |
|---|---|---|---|---|---|---|---|---|---|---|---|
| [21] | 2019 | Journal | Towards an Interoperable Internet of Things Through a Web of virtual things at the Fog layer | Conceptual Research | Framework | Interoperability | 1 | 1 | 1 | 2 | 5 |
| [61] | 2019 | Journal | Contribution of the Web of Things and of the Opportunistic Computing to the Smart Agriculture: A Practical Experiment | Experimental Research | Framework | Monitoring and Controlling | 1 | 1 | 1 | 1.5 | 4.5 |
| [62] | 2014 | Conference | Publishing Danish Agricultural Government Data as Semantic Web Data | Solution Proposal | Method | Publishing | 1 | 1 | 0.5 | 0 | 2.5 |
| [63] | 2020 | Journal | An Application of the A'WOT Analysis for the Management of Cultural Heritage Assets: The Case of the Historical Farmhouses in the Aglié Castle (Turin) | Solution Proposal | Application | Management | 1 | 1 | 1 | 1.5 | 4.5 |
| [64] | 2018 | Conference | Contribution to the Setting up of a Remote Practical Work Platform for STEM: The Case of Agriculture | Contributional Research | Platform | Education | 1 | 1 | 1 | 0.5 | 3.5 |
| [65] | 2018 | Journal | Smart sensors from ground to cloud and web intelligence | Contributional Research | Guideline | Analysis | 1 | 1 | 1 | 1.5 | 4.5 |

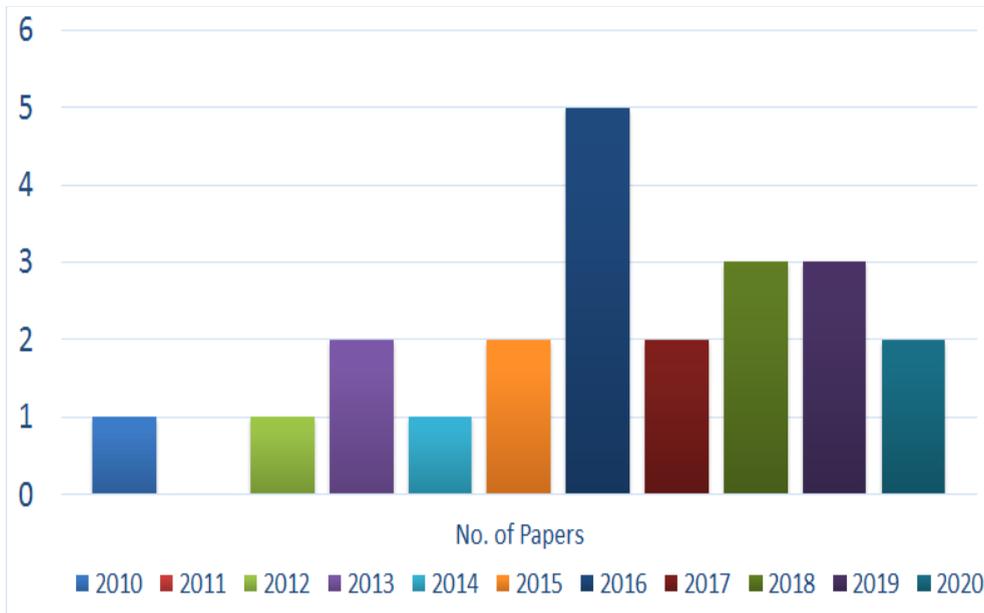

FIGURE 14: Trends in studies by year

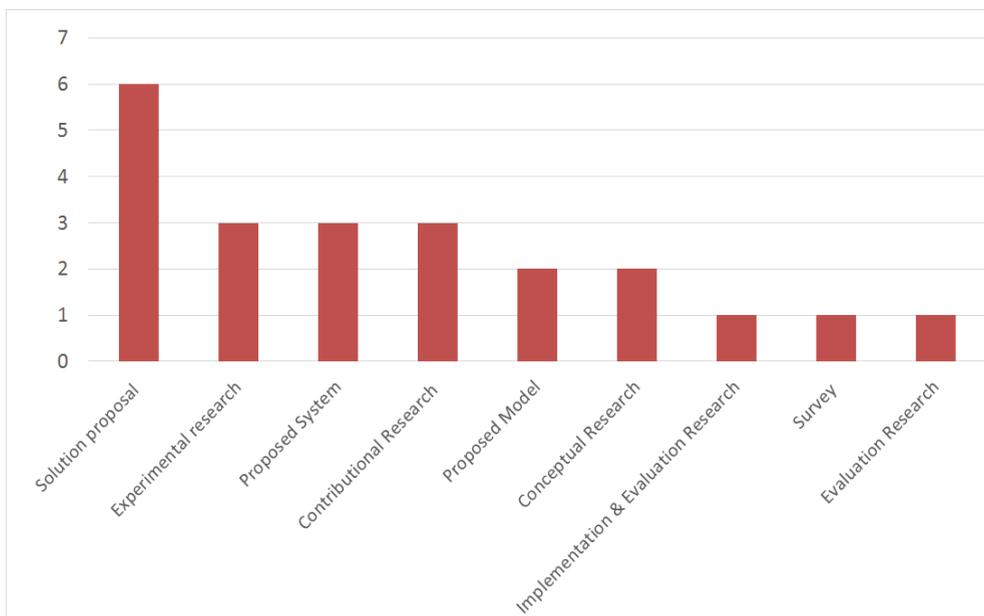

FIGURE 15: Research types

The application consisted of three tabs: Locn tab for showing the locations, Map tab for selecting the desired location, and Data tab for viewing the desired location data. In this article [50] author analyzed the issues regarding the integration of heterogeneous devices and platforms and proposed a framework by using the Semantic web technologies to integrate the multiple domains, devices, platform for streaming and analyzing real-time data, finding the errors, and interoperating the diverse devices, networks, and platforms.

### E. RQ4. WHAT ARE THE CURRENT RESEARCH TOPICS AND GAPS IN WOT AGRICULTURAL RESEARCH?

All the research topics addressed in the elected studies recorded in Table 4. As shown in the results it is concluded that most of the studies are based on the monitoring and management systems, Frameworks, and web applications for the Agricultural domain by integrating the WOT with several other technologies. Some of the cases mentioned here: In this study [53] the author presented a middleware architecture by integrating the Web technology in the existing IoT systems for storing the data of every sensor device with a unique identifier separately in real-time and presented the data in a graphical view.

This study [45] presented a pest forecasting and information system for specified region pest prediction. This web interface-based software used the WWW technologies and provided the storing and mining functions on the data. This system contained the WebGrabber for getting data and then performed the real-time calculation. Then this system enhanced the forecasting and decision-making system by creating an algorithm. This system had the feature of less cost usage, reusable and extendable virtually. A paper [47] proposed a monitoring system for farmers to control and monitor the different situations related to the crops. This system developed by integrating the Wireless sensors networks, Web of Things, and Internet of Things technologies and had the features of an alert system with SMS alert service, real-time monitoring of soil, water irrigation conditions, graphical view, and decision making based on the crop situations.

A study [63] introduced an Application named A'WOT regarding the preservation of historical culture assets. This application could make decisions, managing resources, detecting the problems, and finding the solutions to the problems. Several WOT-based solutions are presented related to the management, monitoring, controlling, and decision making. Most of them integrated the two or three technologies for efficient performance and results but some of them still lack in achieving the goals such as handling huge data amount, multi-domain devices, networks.

Most of the selected studies in this systematic literature review are based on the integration of WOT with IOT-based agricultural systems and applications. To overcome the drawbacks and limitations of IoT-agriculture research, WOT technologies had emerged to get real-time processing and functions in the agricultural domain. WOT services enabled real-time data gathering, analyzing, efficient decision making, efficient crops monitoring, and controlling and handling the heterogeneous devices, sensors, and infrastructures.

But there are still some gaps that were realized in this systematic literature review. The WOT enhanced the standards and infrastructures but there is still a need for secured international standards. Another gap is related to the available agriculture data that should be analyzed before applying it to the actual farms. An important gap has been found regarding the educational and research area of the agricultural domain that it lacks remote practical platforms for educational purposes and only 1 study was found on this topic. Another gap might be seen that there is still less work found on SLR, SR, LR, and Mapping Studies on the WOT-Agricultural domain research area. So it needs further work.

### F. RQ5. WHAT ARE THE APPROACHES THAT WERE PRESENTED TO ADDRESS THE PROBLEMS IN WOT AGRICULTURAL RESEARCH?

In selected studies most of the approaches presented systems (4)(0.2%), Applications (4)(0.2%), Framework (4)(0.2%), guideline (3)(0.14%) and Platform (3)(0.14%). While the left approaches presented Architecture (2)(0.09%), Analysis (1)(0.045%), Method (1)(0.045%) and infrastructure (1)(0.045%). Figure 16 shows the approaches that were presented in the researches. Moreover, The approaches described in the selected studies summarized in the Table 5.

All the approaches were presented to improve the agricultural methods, machines, product qualities for profitable agriculture. Some approaches enhanced the previously presented solutions by involving the wot to make them cost-effective, reliable, and efficient agricultural applications and shift traditional agriculture to smart and precision agriculture.

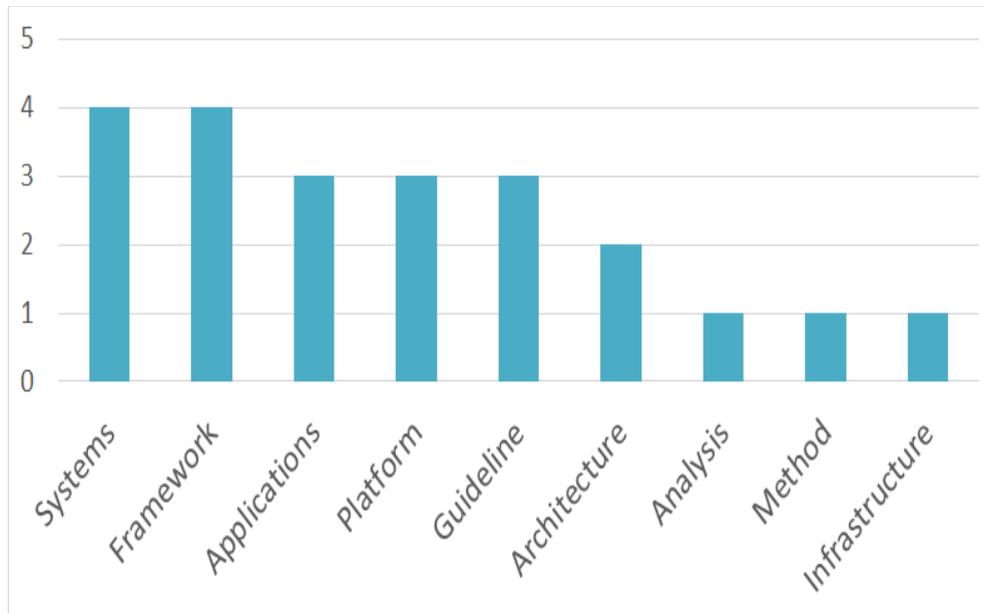

FIGURE 16: Research Approaches

TABLE 5: Approaches Summary of elected Articles

| Study Ref. | Approach |
|---|---|
| [20] | An open geospatial web service-oriented physical infrastructure was proposed to overcome the challenge of interoperability by integrating it in the sensors and systems for PA monitoring. It integrates, processes, and share information from systems and sensors to the world wide web. The ten different sensors were installed in the field of Wuhan city china and conducted experiments to check infrastructure performance. The proposed infrastructure was compared with the existing one that showed the great capability of the proposed one in monitoring processes. |
| [41] | a new precision agriculture management system (PAMS) was developed by integrating two technologies IOT and WebGIS and installed the selected farm. IoT was used for perception accuracy and WebGIS efficiently processed the network geographical information. It contained four stages: Agriculture management platform, spatial information infrastructure platform, mobile client, and IOT infrastructure platform. IoT, WebGIS, Internet and communication, Location-Based Service, GPS, RS techniques were used to collect, transfer and publish the data to help the users in monitoring and managing the production. |
| [42] | This study researched Agricultural Knowledge-Based Systems that used Semantic Technologies to overcome the existing challenges. knowledge-based systems were used by researchers and the general public to understand the information, gaps, and limitations of their related disciplines, provide new ideas and future work, investigate the neglected crops, and utilized their information to make them useful. Many ontologies were developed that helped in the development of semantic products. Many decision and expert systems classified into the knowledge-based systems and web-based systems that used the "if-then" query to provide the answers to the questions. |

TABLE 5: Remaining Approaches Summary of elected Articles

| Study Ref. | Approach |
|---|---|
| [45] | A software was developed in this work that used the World Wide Web technologies for storage and data mining purposes on weather data. It connected the NOA's network stations, downloaded the files with the .txt extension. the software was enhanced by using the www to perform real-time pest forecasting. It has two stages: on the first stage, a customized WebGrabber obtained the weather data, the second one performed the real-time population calculation and stored information on MYSQL local server. An algorithm was developed for decision-making systems and forecasting. This system was easily useable, less cost, and simple architecture that virtually extendable |
| [46] | A web-based architecture was proposed to track, monitor, and analyze the real-time agricultural data. The sensors' data was stored and shared on the Web in the systems. Protocol named "Message Queue Telemetry Transport (MQTT)" was utilized for data transferring and monitoring devices. It contained the features of less power consumption, less complexity level, asynchronous communication. |
| [47] | A Monitoring system using WOT was proposed for effective monitoring and decision making. Wireless Sensor Networks, IoT, and WOT were integrated to efficiently monitor and control crop-related situations. Presented an alarm system for estimating water level, an application to manage the irrigation procedure, a graphical representation, real-time monitoring for soil situations. |
| [48] | Providing the awareness to the farmers about necessary water supply for crops was discussed and then A 1622WQ web-based application was developed and designed as user-centered for providing the real-time information to the farmers. limited internet connection, poor data quality, and operational issues were identified and taken care of in the application. The web-based application was created by integrating the technologies and provided a single data and communication platform to farmers. It also highlighted the significance of collaboration and integration of heterogeneous approaches and technologies for Smart Agriculture. |
| [49] | Research was conducted about improving the productivity of crops and farms by knowing the accurate, necessary weather and environment conditions. So the emergence of IoT and other technologies such as IoT devices, wireless sensors, and their networks, mobiles, smart systems, monitoring cameras, and weather stations was used. Smart-FarmNet the first platform in the world was developed by integrating Semantic Web Technology for automatic collection of data, virtually integrated the diverse devices and sensors, and stored the data to perform analysis. A common API for managing and integrating massive amount of diverse devices and networks for representing their data through SWT was designed. |

TABLE 5: Remaining Approaches Summary of elected Articles

| Study Ref. | Approach |
|---|---|
| [50] | Agri-IoT Framework with the integration of semantic web technologies was developed for enhanced features to support the heterogeneous platforms, devices, sensors, and real-time data streaming in agriculture applications. It provided a common platform with pipeline processing for applications, to analyze the huge data, detected unknown events, and interoperated the heterogeneous devices, sensors, networks. performance of this framework on the medium to large range farms showed great results. |
| [51] | Approach stated that the Applications of WOT required advancement in the solutions to its adaptation in common models for various goals. A solution with the features of reusability and multi-purpose settings was provided. A Smart agriculture framework demonstrates the proposed approach for achieving the accuracy and performance of implementation and evaluation. The outcomes validated the objectives about processing and question-answer adaptation |
| [53] | A specialized Middleware architecture was proposed with the integration of real-time Web technology in existing IoT systems for the urban area Agriculture. The main function was to store the data of every sensor device with a unique separate identifier by the web-based application that also showed graphical visualization. |
| [54] | A survey was intentionally done to encourage future research on SWTs applications for solutions to the problems in agriculture as SWTs played a vital in solving the Agricultural domain problems as a supportive domain. This approach presented a complete review of the already existed SWTs resources, their methodologies, standards of data interchange, and a survey on current SWTs applications. Two approaches to creating the centered semantic resources for agriculture were described. The semantic web technology applications designed for the agricultural field were also discussed and considered in the study. |
| [27] | This approach described the integration of Linked Open Data and Worldwide semantic web technology. Heterogeneous formats and data kinds led to the evolution of data and integration with the World Wide Web as World Wide Semantic Web (WWSW). The Service-oriented techniques paved the way for the development of semantic technology to overcome many challenges related to performance, efficiency, and availability of various services over the internet world. |
| [58] | This study performed analysis on the contribution of the Web of Things in the agricultural domain. Three steps were taken to perform the analysis. 1.desires and estimates of analyst's region, 2. Movements of agricultural machines producers, 3. Status of existed accomplishments in the domain. This study concluded WOT and IOT are the main building blocks of smart agriculture and precision farming and WOT is the main character for agricultural high productivity and profit. |

TABLE 5: Remaining Approaches Summary of elected Articles

| Study Ref. | Approach |
|---|---|
| [59] | Semantic Web-based knowledge model for gathering, mining, integrating the data from various sources to use for understanding and decision purposes in the agricultural field. Web technologies were used to connect to the ontologies with field-related data and metadata for data extraction. This research work was an addition to the existed work of providing the information to the farmers to make a better decision for huge production with less equipment and resources to fulfill the future needs. |
| [60] | Introduced the WOT model and integration of it with the supply chain process to attain the sharing knowledge facility and to make supply chain effective over networks through RFID for maximizing the agricultural production. |
| [21] | This study reviewed the several IoT infrastructures, different companies' projects, and applications of IoT and presented a conceptual framework related to the integration of IoT and WOT technologies. Contained a server named Web of Virtual Things (WoVT) as a Fog Layer to solve compatibility problems, an interface REST for devices to incorporate at the lowest perception layer. The simulation process was performed that showed the emergence of WOT with IoT can be highly useful and successful. |
| [61] | This described the contribution of WOT in agricultural domain. The integration of IoT into the sensors and devices for environment monitoring enhanced the agriculture field by introducing the decision-taking feature and turned it into intelligent agriculture. |
| [62] | this study proposed a method of gathering and proceeding the data and connecting with different databases by using a web framework named "RDF" for the Danish government in the Danish language. The proposed model procedure gets the data, performs analysis, cleans the data, and then linked it with other data sources. The experiment was also conducted that showed good results. |
| [63] | A'WOT application was presented regarding the preservation of historical culture assets in the Agli, e Castle (Turin). It could make decisions, managing resources, detecting the problems, and finding the solutions to the problems. |
| [64] | It addressed the necessity of a virtual practical platform for STEM studies and took part in the enhancement of distance learning to provide a virtual visiting place and sharing different facilities by joining the features of WOT and WebRTC a multimedia server technologies. By implying this framework a communication among different users could be enabled on a simple web browser and get the data from already deployed devices. |
| [65] | This focused on Embedded Intelligence (EI) an integrating research field. The previous research record, properties, applications, frameworks, and EI research-related gaps were reviewed in this study and demonstrated an intelligent water supply application. |

## G. QUALITY ASSESSMENT

The total of every elected article standard has shown within the Table 6. Mostly elected articles hold a high rating that is approximately 50%, 45.5% hold average score while 4.5% contain the lowest ranking that is shown in figure 17. The attained judgment may be helpful to WOT-Agricultural analyzers and professionals to choose related studies.

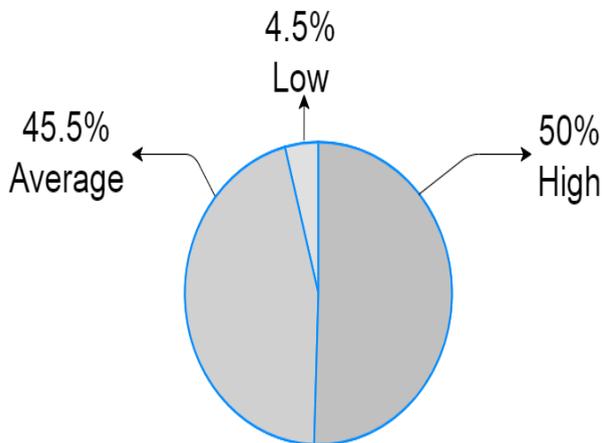

FIGURE 17: Article Ranking

TABLE 6: Quality assessment score

| Ref. | T.score | No. |
|---|---|---|
| [50] | 2 | 1 |
| [62] | 2.5 | 1 |
| [41] [42] [45] [47] [53] [59] [60] | 3 | 7 |
| [51] [64] | 3.5 | 2 |
| [46] [58] | 4 | 2 |
| [48] [49] [54] [27] [61] [63] [65] | 4.5 | 7 |
| [20] [21] | 5 | 2 |

## V. DISCUSSION

This SLR gathered the studies conducted on the WOT agriculture research trends. The selected studies have been used to answer the research questions presented in the methodology section of this study. The results show that many studies have been conducted to find the solutions for agricultural domain problems.

### A. PROPOSED TAXONOMY

The main motive for conducting this systematic literature review was to investigate the continuing research movement of the agricultural technology era by filtering out the 22 studies from 220 studies. After filtering out these papers according to the given rules. The dominant outcomes are given: The review of selected studies conducted on WOT involvement in agriculture recognized several application domains for agriculture research trends. The web of things paved the new path for the agriculture sector to smart agriculture by providing the solutions to problems faced by the IoT agriculture devices, architectures, integrated systems, and much more. Web of things resolved the issues of heterogeneous data, devices, standards, and platforms. These application domains encourage the readers for further research on WOT-agricultural research. The application domains have been classified into eleven types: Monitoring, Controlling, Managements, Storing, Analysis, Optimization, Precision and Decision Making, Publishing, Integration, Education, and Interoperability. These domains are further divided into subdomains such as air, water, weather, devices, systems monitoring and controlling, management of farms sensors, devices, assets, storing the data related to the crops, diseases, researches, results, weather conditions, and perform the mining on it, Analyzing the agricultural data, production, and systems performances, Optimizing the systems and supply chain processes, made the prediction based on the available farms' data and provide the decisions support. Publishing the datasets, experimental results, Integrating the networks, devices, standards, and protocols for better performance.

Providing the study platforms with real-time information, researches implementation results to the learners, and interoperating the diverse IoT devices, networks, sensors for effective smart Agriculture. These applications became possible with the emergence of the Web of things with the IoT and other technologies to overcome the limitations of those technologies. WOT and IOT emergence will help in transforming traditional agriculture to smart agriculture.

The application domains showing in the taxonomy were got from the classification table presented in the research question results section. That classification table was generated by reviewing the 22 selected research papers based on the WOT agricultural existing research trends. These selected papers were evaluated through the inclusion, exclusion criteria presented in the study selection process of methodology section 2. This taxonomy will help in motivating for further research in this area. The proposed taxonomy has shown in the figure 18 below.

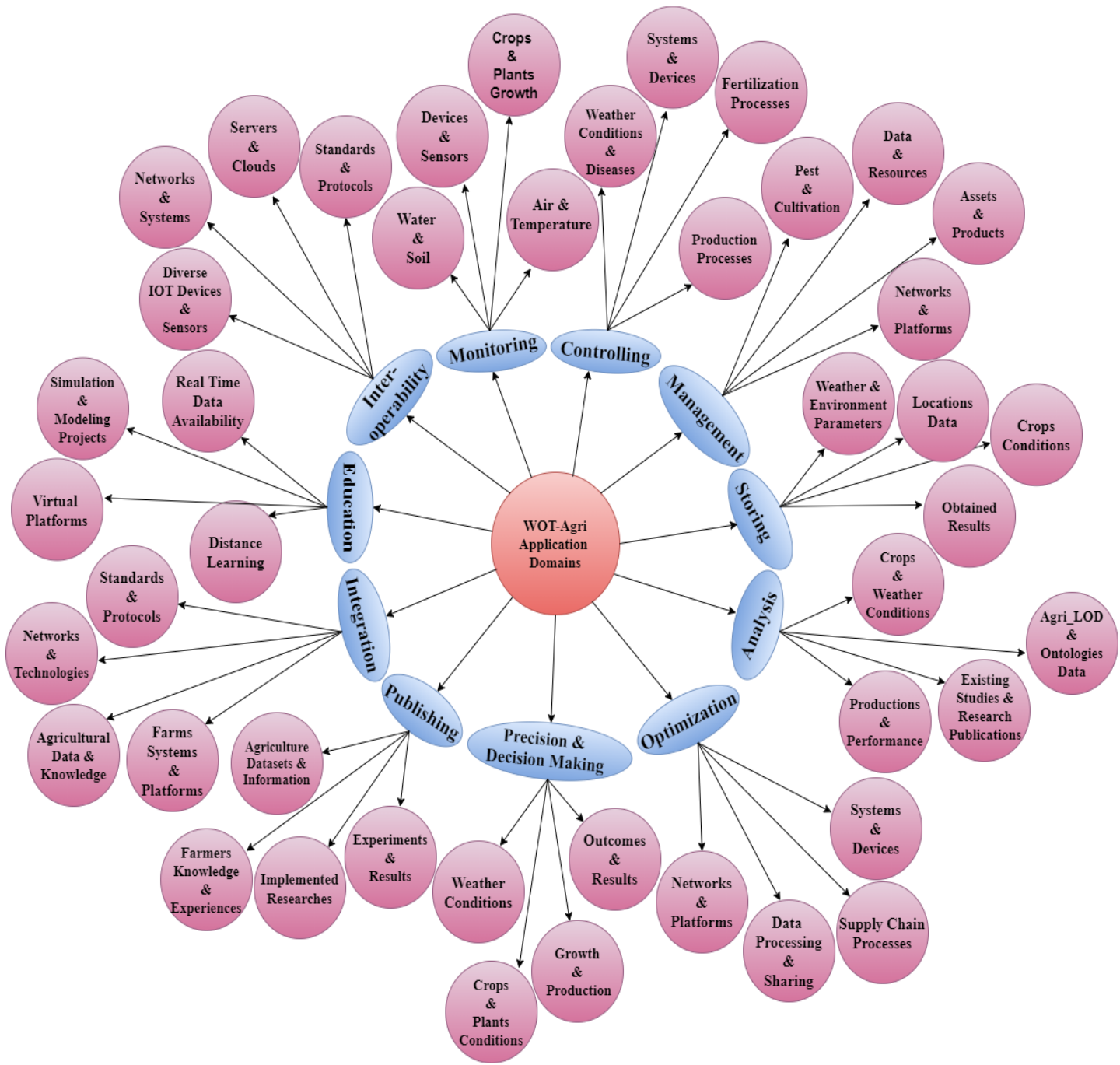

FIGURE 18: Taxonomy of Application Domains

### B. WOT AGRICULTURAL MODEL

A web of things based model has designed to illustrate the WOT contribution in the agriculture to make Smart Agricultural environment. This model explains the application of WOT in the different sectors of Agriculture such as in monitoring, controoling, managing the devices, machines, crops, animals, weather conditions, irrigation systems through web services. One of the most important feature is the enabling of real time services and secure connection. This model presented the picture of WOT based Smart Agriculture in the given figure 20.

### C. OPEN ISSUES AND CHALLENGES

Many studies acquired the wot integration, emergence into the existing presented solution, and in new solutions to provide the solution to the existing agricultural applications and systems problems. WOT became the main actor for addressing the agricultural domain problems. However, there still open challenges and issues about WOT collaboration with agriculture application domains that has shown in the figure 19.

The most prominent are security issues regarding the emergence of different technologies at different layers in the architectures, platforms, infrastructures, web applications. Threats to the reliability of agricultural data, Cost issues on integrating the WOT technologies in hardware and software. Need of international standards for effective security service.

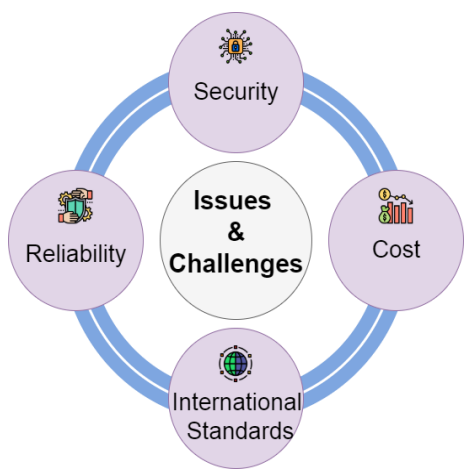

FIGURE 19: Main Issues and Challenges

Awareness to the farmers on adopting the wot applications for their effective farming and production profit. Availability of more literature review, mapping studies, SLR on the adoption, implementation, and emergence of WOT in the agricultural technology sector. Poor knowledge about technologies of the farmers of rural regions. The enhancement and improvement in WOT agriculture applications for the broad level farms and fields. However, there is a need for motivation for further research on existing and ongoing topics.

### D. RESEARCH GAP AND FUTURE DIRECTIONS

The discussion concluded that many studies had been presented for solving the agricultural applications problems by Web of things technologies involvement. Hence, there is still a need for more well-systematic studies on WOT solutions for agriculture domain problems. As previously discussed that the WOT proved itself as the main character for the development of the agricultural sector. Most of the researches presented the existing and new solutions and studies that presented the solutions, without any systematic way. Therefore, there is a need for systematic studies, literature reviews on the existing solutions researches so that it could help in further research and finding the gaps in the current and ongoing researches and solutions to them. Furthermore, more focus is needed on evaluation researches to be conducted for evaluating the existing approaches work related to the WOT trends in the agricultural sector to lead it to smart automating agriculture.

## VI. THREAT TO VALIDITY

This SLR may have some threats to its validity that include an incomplete selection of research articles, imperfect collection of data, and proper quality assessment of the selected studies.

### A. SELECTION OF RESEARCH ARTICLES

A detailed guideline for the selection of the research articles is defined in Section III. The Study Selection Process subsection presented the complete Inclusion/Exclusion criteria for selecting the research articles. The elected studies' years range from 2010 to 2020 for obtaining the ongoing movement of WOT in the agriculture field. Although there remains a possibility of some papers being missed. The primary reason for this possibility is the exclusion of articles that were published before 2010 and general focus articles. The second possibility could be the search string for gathering the research articles. Even though the detailed search string is discussed in Section III, but there is still left a possibility that few research papers might require some other searching keywords. The search string provided in section III is refined several times to find the most relevant research articles, but there is still lie down a gap for the new keyword.

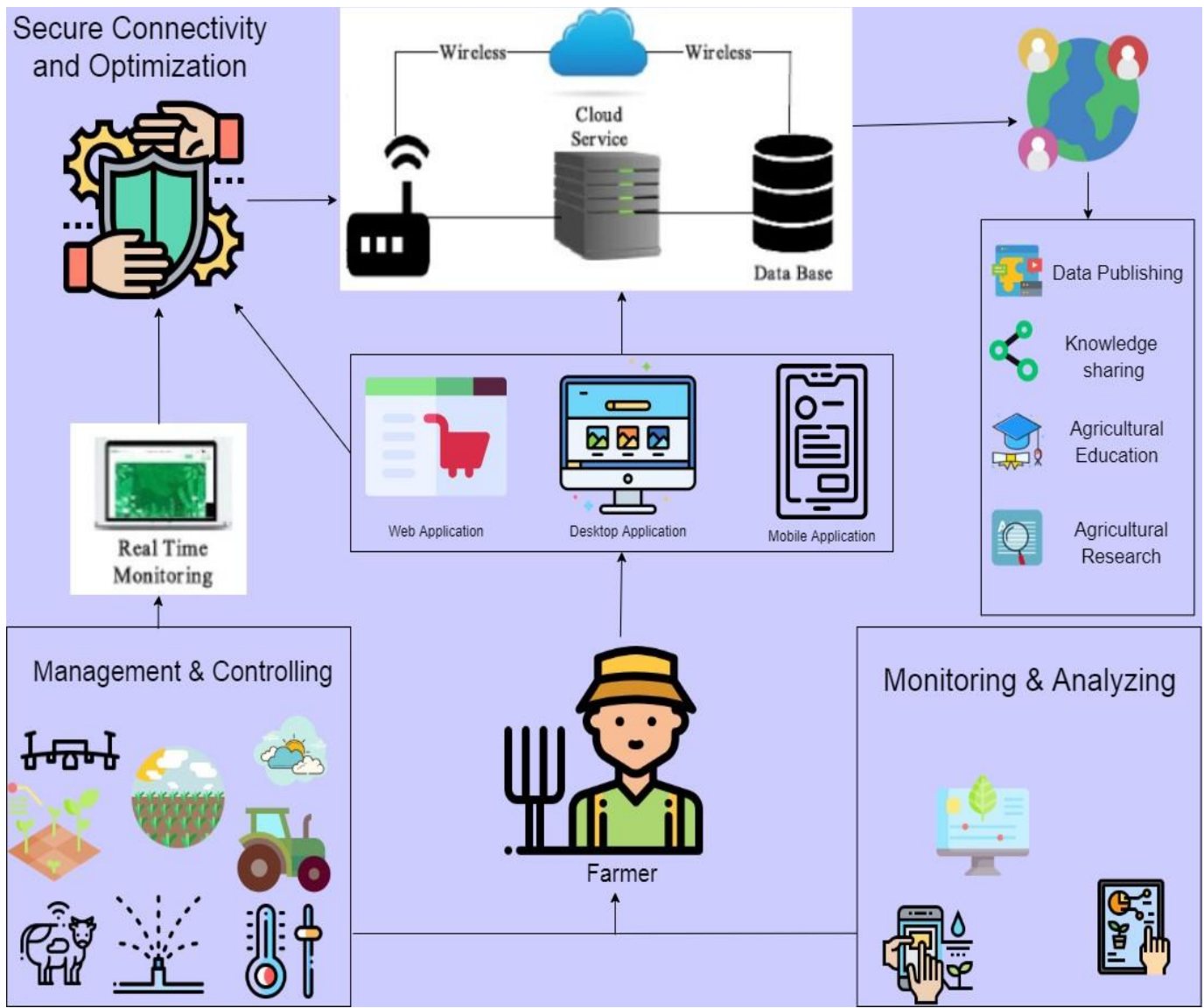

FIGURE 20: WOT Agricultural Model

### B. INACCURATE COLLECTION OF DATA

Another realizable reason for indefinite results is the incomplete collection of data. This possibility has been decreased by reviewing the gathered data three times.

### C. QUALITY EVALUATION

One of the crucial tasks in SLR is the quality evaluation because the unreliable quality assessment could be led to unreliable conclusions. A suitable procedure to attain the quality of the selected research articles has been taken out and discussed in Section IV.

## VII. CONCLUSION

This study describes a systematic literature review of existing researches published on the Web of Things-based agriculture domain. This comprehensive survey has been conducted based on 22 studies selected according to the well-defined methodology. The review was done according to the defined rules carefully and filtered out the main works. Each and every work was thoroughly checked and recognized for effective outputs. After doing a thorough review of previous studies, it concluded that the WOT technology introduced promising solutions to the agriculture applications domain problems. WOT amazingly overcomes the interoperability drawbacks of IoT agriculture devices and systems by providing web-enabled services, improved the old systems, helped in making online informative databases for farmers, educational institutes as well as the general public.

Considering this situation, we have presented a taxonomy of agriculture application domains where the web of things provided the services to overcome the existing technological challenges in the agriculture domain. It expresses the areas of agriculture where wot services were used in various ways such as in new and old schemes for reducing the complexities, wide range flaws. A model of wot based smart agriculture described the working in the agricultural field by integrating the web of things. This taxonomy and model will help the researchers in getting a clear picture of agricultural domains where WOT was used and is being used, understanding them, and selecting the desired application domains according to their needs. It will indicate the paths for more new domains where WOT can be useful in handling the farming and other agriculture processes.

The promising future directions in this study involve the identification of how these application domains are helping in motivating the readers. Furthermore, another important fact is that the WOT-based agriculture research is being supported by the governments of several countries as well as many countries that have their WOT agriculture policies. Although, many research works have been conducted on WOT-based agriculture not much of the work presented as SLR, SMS, and SR. Thus, more studies need to be conducted in this area to motivate further research.

### REFERENCES

[1] I. Mat, M. R. Mohd Kassim, A. N. Harun, and I. M. Yusoff, "Smart agriculture using internet of things," in 2018 IEEE Conference on Open Systems (ICOS), pp. 54–59, 2018.

[2] I. Charania and X. Li, "Smart farming: Agriculture's shift from a labor intensive to technology native industry," Internet of Things, vol. 9, p. 100142, 2020.

[3] J. Lindblom, C. Lundström, M. Ljung, and A. Jonsson, "Promoting sustainable intensification in precision agriculture: review of decision support systems development and strategies," Precision Agriculture, vol. 18, no. 3, pp. 309–331, 2017.

[4] A. Balafoutis, B. Beck, S. Fountas, J. Vangeyte, T. V. d. Wal, I. Soto, M. Gómez-Barbero, A. Barnes, and V. Eory, "Precision agriculture technologies positively contributing to ghg emissions mitigation, farm productivity and economics," Sustainability, vol. 9, no. 8, 2017.

[5] A. P. Barnes, I. Soto, V. Eory, B. Beck, A. Balafoutis, B. Sánchez, J. Vangeyte, S. Fountas, T. van der Wal, and M. Gómez-Barbero, "Influencing factors and incentives on the intention to adopt precision agricultural technologies within arable farming systems," Environmental Science and Policy, vol. 93, pp. 66–74, 2018.

[6] A. Knierim, M. Kernecker, K. Erdle, T. Kraus, F. Borges, and A. Wurbs, "Smart farming technology innovations – insights and reflections from the german smart-akis hub," NJAS - Wageningen Journal of Life Sciences, vol. 90-91, p. 100314, 2019.

[7] A. Marucci, A. Colantoni, I. Zambon, and G. Egidi, "Precision farming in hilly areas: The use of network rtk in gnss technology," Agriculture, vol. 7, no. 7, 2017.

[8] A. T. Balafoutis, B. Beck, S. Fountas, Z. Tsiropoulos, J. Vangeyte, T. van der Wal, I. Soto-Embodas, M. Gómez-Barbero, and S. M. Pedersen, "Smart farming technologies–description, taxonomy and economic impact," in Precision Agriculture: Technology and Economic Perspectives, pp. 21–77, Springer, 2017.

[9] M. S. Farooq, S. Riaz, A. Abid, T. Umer, and Y. B. Zikria, "Role of iot technology in agriculture: A systematic literature review," Electronics, vol. 9, no. 2, 2020.

[10] D. Evans, "The internet of things: How the next evolution of the internet is changing everything," CISCO white paper, vol. 1, no. 2011, pp. 1–11, 2011.

[11] C. V. N. Index, "Forecast and methodology, 2012–2017," White paper, vol. 29, 2013.

[12] A. Mazayev, J. A. Martins, and N. Correia, "Semantic web thing architecture," in 2017 4th Experiment@ International Conference (exp. at'17), pp. 43–46, IEEE, 2017.

[13] D. Guinard, V. Trifa, F. Mattern, and E. Wilde, "From the internet of things to the web of things: Resource-oriented architecture and best practices," in Architecting the Internet of things, pp. 97–129, Springer, 2011.

[14] D. Raggett, "The web of things: Challenges and opportunities," Computer, vol. 48, no. 5, pp. 26–32, 2015.

[15] E. Antonopoulou, S. Karetsos, M. Maliappis, and A. Sideridis, "Web and mobile technologies in a prototype dss for major field crops," Computers and Electronics in Agriculture, vol. 70, no. 2, pp. 292–301, 2010.

[16] D. L. Hernández-Rojas, T. M. Fernández-Caramés, P. Fraga-Lamas, and C. J. Escudero, "A plug-and-play human-centered virtual teds architecture for the web of things," Sensors, vol. 18, no. 7, p. 2052, 2018.

[17] M. Khan, B. N. Silva, and K. Han, "A web of things-based emerging sensor network architecture for smart control systems," Sensors, vol. 17, no. 2, 2017.

[18] A. Rhayem, M. B. A. Mhiri, and F. Gargouri, "Semantic web technologies for the internet of things: Systematic literature review," Internet of Things, p. 100206, 2020.

[19] F. Aznoli and N. J. Navimipour, "Deployment strategies in the wireless sensor networks: systematic literature review, classification, and current trends," Wireless Personal Communications, vol. 95, no. 2, pp. 819–846, 2017.

[20] N. Chen, X. Zhang, and C. Wang, "Integrated open geospatial web service enabled cyber-physical information infrastructure for precision agriculture monitoring," Computers and Electronics in Agriculture, vol. 111, pp. 78–91, 2015.

[21] B. Negash, T. Westerlund, and H. Tenhunen, "Towards an interoperable internet of things through a web of virtual things at the fog layer," Future Generation Computer Systems, vol. 91, pp. 96–107, 2019.

[22] F. Bauer and M. Kaltenböck, "Linked open data: The essentials," Edition mono/monochrom, Vienna, vol. 710, 2011.

[23] T. Katayama, M. D. Wilkinson, G. Micklem, S. Kawashima, A. Yamaguchi, M. Nakao, Y. Yamamoto, S. Okamoto, K. Oouchida, H.-W. Chun, et al., "The 3rd dbcls biohackathon: improving life science data integration with semantic web technologies," Journal of biomedical semantics, vol. 4, no. 1, pp. 1–17, 2013.

[24] T. Heath and C. Bizer, "Linked data: Evolving the web into a global data space," Synthesis lectures on the semantic web: theory and technology, vol. 1, no. 1, pp. 1–136, 2011.

[25] O. Aziz, T. Anees, and E. Mehmood, "An efficient data access approach with queue and stack in optimized hybrid join," IEEE Access, pp. 1–1, 2021.

[26] H.-G. Kim, "Semantic web," 2003.

[27] D. Lukose et al., "World-wide semantic web of agriculture knowledge," Journal of Integrative Agriculture, vol. 11, no. 5, pp. 769–774, 2012.

[28] C. Caracciolo, A. Morshed, A. Stellato, G. Johannsen, Y. Jaques, and J. Keizer, "Thesaurus maintenance, alignment and publication as linked data: The agroovoc use case," in Metadata and Semantic Research (E. García-Barriocanal, Z. Cebeci, M. C. Okur, and A. Öztürk, eds.), (Berlin, Heidelberg), pp. 489–499, Springer Berlin Heidelberg, 2011.

[29] A. Chehri, H. Chaibi, R. Saadane, N. Hakem, and M. Wahbi, "A framework of optimizing the deployment of iot for precision agriculture industry," Procedia Computer Science, vol. 176, pp. 2414–2422, 2020.

[30] S. Wolfert, L. Ge, C. Verdouw, and M.-J. Bogaardt, "Big data in smart farming – a review," Agricultural Systems, vol. 153, pp. 69–80, 2017.

[31] A. R. de Araujo Zanella, E. da Silva, and L. C. P. Albini, "Security challenges to smart agriculture: Current state, key issues, and future directions," Array, p. 100048, 2020.

[32] H. S. Abdullahi, F. Mahieddine, and R. E. Sheriff, "Technology impact on agricultural productivity: A review of precision agriculture using unmanned aerial vehicles," in Wireless and Satellite Systems (P. Pillai, Y. F. Hu, I. Otung, and G. Giambene, eds.), (Cham), pp. 388–400, Springer International Publishing, 2015.

[33] F. S. Khan, S. Razzaq, K. Irfan, F. Maqbool, A. Farid, I. Illahi, and T. U. Amin, "Dr. wheat: a web-based expert system for diagnosis of diseases and pests in pakistani wheat," in Proceedings of the World Congress on Engineering, vol. 1, pp. 2–4, Citeseer, 2008.

[34] A. Walter, R. Finger, R. Huber, and N. Buchmann, "Opinion: Smart farming is key to developing sustainable agriculture," Proceedings of the National Academy of Sciences, vol. 114, no. 24, pp. 6148–6150, 2017.

[35] A. D. Boursianis, M. S. Papadopoulou, P. Diamantoulakis, A. Liopa-Tsakalidi, P. Barouchas, G. Salahas, G. Karagiannidis, S. Wan, and S. K. Goudos, "Internet of things (iot) and agricultural unmanned aerial vehicles (uavs) in smart farming: A comprehensive review," Internet of Things, p. 100187, 2020.

[36] M. Torky and A. E. Hassanein, "Integrating blockchain and the internet of things in precision agriculture: Analysis, opportunities, and challenges," Computers and Electronics in Agriculture, p. 105476, 2020.

[37] A. Kamilaris, A. Fonts, and F. X. Prenafeta-Bold, "The rise of blockchain technology in agriculture and food supply chains," Trends in Food Science Technology, vol. 91, pp. 640–652, 2019.

[38] D. Pivoto, P. D. Waquil, E. Talamini, C. P. S. Finocchio, V. F. Dalla Corte, and G. de Vargas Mores, "Scientific development of smart farming technologies and their application in brazil," Information processing in agriculture, vol. 5, no. 1, pp. 21–32, 2018.

[39] D. Glaroudis, A. Iossifides, and P. Chatzimisios, "Survey, comparison and research challenges of iot application protocols for smart farming," Computer Networks, vol. 168, p. 107037, 2020.

[40] S. Blank, C. Bartolein, A. Meyer, R. Ostermeier, and O. Rostanin, "igreen: A ubiquitous dynamic network to enable manufacturer independent data exchange in future precision farming," Computers and electronics in agriculture, vol. 98, pp. 109–116, 2013.

[41] J. Ye, B. Chen, Q. Liu, and Y. Fang, "A precision agriculture management system based on internet of things and webgis," in 2013 21st International Conference on Geoinformatics, pp. 1–5, IEEE, 2013.

[42] E. Jahanshiri and S. Walker, "Agricultural knowledge-based systems at the age of semantic technologies," Inter. J. Know. Engin, vol. 1, no. 1, pp. 64–67, 2015.

[43] Wikipedia contributors, "Knowledge-based systems — Wikipedia, the free encyclopedia," 2021. [Online; accessed 16-April-2021].

[44] C. Yialouris and A. Sideridis, "An expert system for tomato diseases," Computers and electronics in agriculture, vol. 14, no. 1, pp. 61–76, 1996.

[45] P. Damos, S. Karabatakis, et al., "Real time pest modeling through the world wide web: decision making from theory to praxis," IOBC-WPRS Bulletin, vol. 91, pp. 253–258, 2013.

[46] K. Grgić, I. Špeh, and I. Heđi, "A web-based iot solution for monitoring data using mqtt protocol," in 2016 international conference on smart systems and technologies (SST), pp. 249–253, IEEE, 2016.

[47] F. Karim, F. Karim, et al., "Monitoring system using web of things in precision agriculture," Procedia Computer Science, vol. 110, pp. 402–409, 2017.

[48] M. P. Vilas, P. J. Thorburn, S. Fielke, T. Webster, M. Mooij, J. S. Biggs, Y.-F. Zhang, A. Adham, A. Davis, B. Dungan, et al., "1622wq: A web-based application to increase farmer awareness of the impact of agriculture on water quality," Environmental Modelling & Software, vol. 132, p. 104816, 2020.

[49] P. P. Jayaraman, A. Yavari, D. Georgakopoulos, A. Morshed, and A. Zaslavsky, "Internet of things platform for smart farming: Experiences and lessons learnt," Sensors, vol. 16, no. 11, p. 1884, 2016.

[50] A. Kamilaris, F. Gao, F. X. Prenafeta-Boldu, and M. I. Ali, "Agri-iot: A semantic framework for internet of things-enabled smart farming applications," in 2016 IEEE 3rd World Forum on Internet of Things (WF-IoT), pp. 442–447, IEEE, 2016.

[51] M. Terdjimi, L. Médini, M. Mrissa, and M. Maleshkova, "Multi-purpose adaptation in the web of things," in International and Interdisciplinary Conference on Modeling and Using Context, pp. 213-226, Springer, 2017.

[52] M. Mrissa, L. Médini, J.-P. Jamont, N. Le Sommer, and J. Laplace, "An avatar architecture for the web of things," IEEE Internet Computing, vol. 19, no. 2, pp. 30–38, 2015.

[53] A. Ordoñez-García, E. V. Núñez, M. Siller, and M. G. S. Cervantes, "Iot system for agriculture: Web technologies in real time with the middleware paradigm.," in 2018 IEEE International Autumn Meeting on Power, Electronics and Computing (ROPEC), pp. 1–4, IEEE, 2018.

[54] B. Drury, R. Fernandes, M.-F. Moura, and A. de Andrade Lopes, "A survey of semantic web technology for agriculture," Information Processing in Agriculture, vol. 6, no. 4, pp. 487–501, 2019.

[55] S. Staab, "Ontology engineering," 2009.

[56] W. Qian, T. Lan, and Z. Lijun, "Approach to ontology construction based on text mining," New Zealand Journal of Agricultural Research, vol. 50, no. 5, pp. 1383–1391, 2007.

[57] F. Amarger, J.-P. Chanet, O. Haemmerlé, N. Hernandez, and C. Roussey, "Skos sources transformations for ontology engineering: Agronomical taxonomy use case," in Research Conference on Metadata and Semantics Research, pp. 314–328, Springer, 2014.

[58] G. Donca et al., "Aspects of wot contribution to sustainable agricultural production.," Analele Universității din Oradea, Fascicula: Protecția Mediului, vol. 26, pp. 27–34, 2016.

[59] M. Shoaib and A. Basharat, "Semantic web based integrated agriculture

[59] information framework," in 2010 Second International Conference on Computer Research and Development, pp. 285–289, IEEE, 2010.

[60] L. Liu and D. Ling, "Discussion on the optimization of web of things supply chain of agricultural products and information sharing based on rfid," in Journal of Physics: Conference Series, vol. 1578, p. 012114, IOP Publishing, 2020.

[61] L. Touseau and N. Le Sommer, "Contribution of the web of things and of the opportunistic computing to the smart agriculture: A practical experiment," Future Internet, vol. 11, no. 2, p. 33, 2019.

[62] A. B. Andersen, N. Gür, K. Hose, K. A. Jakobsen, and T. B. Pedersen, "Publishing danish agricultural government data as semantic web data," in Joint International Semantic Technology Conference, pp. 178–186, Springer, 2014.

[63] M. Bottero, C. D'Alpaos, and A. Marello, "An application of the a'wot analysis for the management of cultural heritage assets: The case of the historical farmhouses in the aglié castl (turin)," Sustainability, vol. 12, no. 3, p. 1071, 2020.

[64] B. M. Degboe, U. H. S. Boko, K. Gueye, and S. Ouya, "Contribution to the setting up of a remote practical work platform for stem: The case of agriculture," in International Conference on e-Infrastructure and e-Services for Developing Countries, pp. 88–97, Springer, 2018.

[65] W. Yong, L. Shuaishuai, L. Li, L. Minzan, L. Ming, K. Arvanitis, C. Georgieva, and N. Sigrimis, "Smart sensors from ground to cloud and web intelligence," IFAC-PapersOnLine, vol. 51, no. 17, pp. 31–38, 2018.

[66] O. Aziz, M. S. Farooq, A. Abid, R. Saher, and N. Aslam, "Research trends in enterprise service bus (esb) applications: a systematic mapping study," IEEE Access, vol. 8, pp. 31180–31197, 2020.

[67] I. Obaid, M. S. Farooq, and A. Abid, "Gamification for recruitment and job training: Model, taxonomy, and challenges," IEEE Access, vol. 8, pp. 65164–65178, 2020.

[68] M. S. Farooq, S. Riaz, A. Abid, T. Umer, and Y. B. Zikria, "Role of iot technology in agriculture: A systematic literature review," Electronics, vol. 9, no. 2, p. 319, 2020.

[69] Z. A. Barmi, A. H. Ebrahimi, and R. Feldt, "Alignment of requirements specification and testing: A systematic mapping study," in 2011 IEEE Fourth International Conference on Software Testing, Verification and Validation Workshops, pp. 476–485, IEEE, 2011.

[70] E. Jahanshiri and S. Walker, "Agricultural knowledge-based systems at the age of semantic technologies," International Journal of Knowledge Engineering, vol. 1, pp. 64–67, 01 2015.

[71] P. Damos and S. Karabatak'Is, "Real time pest modeling through the world wide web: decision making from theory to praxis.," IOBC/WPRS Bulletin, vol. 91, pp. 253–258, 2013.

[72] K. Grgić, I. Špeh, and I. Heđi, "A web-based iot solution for monitoring data using mqtt protocol," in 2016 International Conference on Smart Systems and Technologies (SST), pp. 249–253, 2016.

[73] F. Karim, F. Karim, and A. frihida, "Monitoring system using web of things in precision agriculture," Procedia Computer Science, vol. 110, pp. 402–409, 2017. 14th International Conference on Mobile Systems and Pervasive Computing (MobiSPC 2017) / 12th International Conference on Future Networks and Communications (FNC 2017) / Affiliated Workshops.

[74] M. P. Vilas and P. J, "1622wq: A web-based application to increase farmer awareness of the impact of agriculture on water quality," Environmental Modelling Software, vol. 132, p. 104816, 2020.

[75] P. P. Jayaraman, A. Yavari, D. Georgakopoulos, A. Morshed, and A. Zaslavsky, "Internet of things platform for smart farming: Experiences and lessons learnt," Sensors, vol. 16, no. 11, 2016.

[76] A. Kamilaris, F. Gao, F. X. Prenafeta-Boldu, and M. I. Ali, "Agri-iot: A semantic framework for internet of things-enabled smart farming applications," in 2016 IEEE 3rd World Forum on Internet of Things (WF-IoT), pp. 442–447, 2016.

[77] M. Terdjimi, L. Médini, M. Mrissa, and M. Maleshkova, "Multi-purpose adaptation in the web of things," in Modeling and Using Context (P. Brézillon, R. Turner, and C. Penco, eds.), (Cham), pp. 213–226, Springer International Publishing, 2017.

[78] A. Ordoñez-García, E. V. Núñez, M. Siller, and M. G. S. Cervantes, "Iot system for agriculture: Web technologies in real time with the middleware paradigm.," in 2018 IEEE International Autumn Meeting on Power, Electronics and Computing (ROPEC), pp. 1–4, 2018.

[79] B. Drury, R. Fernandes, M.-F. Moura, and A. de Andrade Lopes, "A survey of semantic web technology for agriculture," Information Processing in Agriculture, vol. 6, no. 4, pp. 487–501, 2019.

[80] G. Donca, "Aspects of wot contribution to sustainable agricultural production.," Analele Universității din Oradea, Fascicula: Protecția Mediului, vol. 26, pp. 27–34, 2016.

[81] M. Shoaib and A. Basharat, "Semantic web based integrated agriculture information framework," in 2010 Second International Conference on Computer Research and Development, pp. 285–289, 2010.

[82] L. Liu and D. Ling, "Discussion on the optimization of web of things supply chain of agricultural products and information sharing based on RFID," Journal of Physics: Conference Series, vol. 1578, p. 012114, jul 2020.

[83] B. Negash, T. Westerlund, and H. Tenhunen, "Towards an interoperable internet of things through a web of virtual things at the fog layer," Future Generation Computer Systems, vol. 91, pp. 96–107, 2019.

[84] L. Touseau and N. Le Sommer, "Contribution of the web of things and of the opportunistic computing to the smart agriculture: A practical experiment," Future Internet, vol. 11, no. 2, 2019.

[85] A. B. Andersen, N. Gür, K. Hose, K. A. Jakobsen, and T. B. Pedersen, "Publishing danish agricultural government data as semantic web data," in Semantic Technology (T. Supnithi, T. Yamaguchi, J. Z. Pan, V. Wuwongse, and M. Buranarach, eds.), (Cham), pp. 178–186, Springer International Publishing, 2015.

[86] M. Bottero, C. D'Alpaos, and A. Marello, "An application of the a'wot analysis for the management of cultural heritage assets: The case of the historical farmhouses in the aglié castle (turin)," Sustainability, vol. 12, no. 3, 2020.

[87] B. M. Degboe, U. H. S. Boko, K. Gueye, and S. Ouya, "Contribution to the setting up of a remote practical work platform for stem: The case of agriculture," in e-Infrastructure and e-Services for Developing Countries (G. Mendy, S. Ouya, I. Dioum, and O. Thiaré, eds.), (Cham), pp. 88–97, Springer International Publishing, 2019.

[88] W. Yong, L. Shuaishuai, L. Li, L. Minzan, L. Ming, K. Arvanitis, C. Georgieva, and N. Sigrimis, "Smart sensors from ground to cloud and web intelligence," IFAC-PapersOnLine, vol. 51, no. 17, pp. 31–38, 2018. 6th IFAC Conference on Bio-Robotics BIOROBOTICS 2018.

...